\documentclass[10pt,twocolumn,amsmath,amssymb,aps,prb]{revtex4-2}

\usepackage{graphicx}
\usepackage{dcolumn}
\usepackage{bm}
\usepackage{hyperref}
\usepackage[cp1250]{inputenc}
\usepackage[usenames,dvipsnames]{color}
\usepackage{tabularx}

\usepackage{cancel}
\usepackage{color}

\newcommand{\rmi}{{\rm i}}

\def\XXint#1#2#3{{\setbox0=\hbox{$#1{#2#3}{\int}$}
     \vcenter{\hbox{$#2#3$}}\kern-.5\wd0}}

\begin{document}

\title{Generic Mott-Hubbard phase diagram
for extended Hubbard models\\ without Umklapp scattering}
  
\author{Florian Gebhard$^1$}
\email{florian.gebhard@physik.uni-marburg.de}
\author{Kevin Bauerbach}
\affiliation{$^1$Fachbereich Physik, Philipps-Universit\"at Marburg,
  35032 Marburg, Germany}
\author{\"Ors Legeza$^{1,2,3}$}
\email{legeza.ors@wigner.hu}
\affiliation{$^2$Strongly Correlated Systems Lend\"ulet Research Group, 
Institute for Solid State Physics and Optics, MTA Wigner Research Centre for
Physics, P.O.\ Box 49, 1525 Budapest, Hungary}
\affiliation{$^3$Institute for Advanced Study,
  Technical University of Munich, Lichtenbergstra\ss e 2a,
  85748 Garching, Germany}

\date{September 9, 2023}

\begin{abstract}%

We determine the ground-state phase diagram for the $1/r$-Hubbard model 
 with repulsive nearest-neighbor interaction at half band-filling
 using the density-matrix renormalization group (DMRG)
method. Due to the absence of Umklapp scattering, 
the phase diagram displays finite regions for the three generic phases,
namely, a Luttinger liquid metal for weak interactions, 
a Mott-Hubbard insulator for dominant Hubbard interactions, and 
a charge-density-wave insulator for dominant nearest-neighbor interactions.
Up to moderate interactions strengths, the quantum phase transitions between the metallic and insulating phases
are continuous, i.e., the gap opens continuously as a function of the interaction strength.
We conclude that generic short-range interactions do not change the nature of the Mott transition qualitatively.
\end{abstract}



\maketitle

\section{Overview}
\label{sec:overview}

After a short introduction in Sect.~\ref{subsec:Intro},
we present in Sect.~\ref{subsec:phasediagram} the generic Mott-Hubbard phase diagram 
for extended Hubbard models without Umklapp scattering, the
central result of our work.
The corresponding model and its ground-state properties are
discussed in the remainder of this work, as outlined in Sect.~\ref{subsec:outline}.

\subsection{Introduction}
\label{subsec:Intro}

The Mott transition is one of the long-standing problems in
condensed-matter many-body physics~\cite{Mottbook,Gebhardbook}.
As formalized in the Hubbard model~\cite{Hubbard1963,Gutzwiller1963,Kanamori},
an electronic system with a single-band of width~$W$ and a purely local interaction
of strength~$U$ will be a metal for weak interactions, $W\ll U$, and
an insulator for strong interactions, $U \gg W$. 
As argued by Mott early on~\cite{Mott1949}, there must be a metal-to-insulator 
transition, generically at $U_{\rm c}\approx W$ when the two energy scales are comparable, irrespective of magnetic or charge order.

The quantitative analysis of a quantum phase transition in an interacting many-particle system is notoriously difficult. Concomitantly, analytical solutions are
scarce even for the simplest models and in one spatial dimension~\cite{LiebWu,Gebhardbook,Essler}. 
Numerical approaches in finite dimensions are hampered by finite-size effects so that 
the calculation of ground-state quantities is also best performed for one-dimensional model systems. 
In one dimension, 
the numerical density-matrix renormalization group (DMRG) method provides accurate data for large enough systems with the order of hundred lattice sites and 
particles~\cite{White-1992a,White-1992b,White-1993,Schollwock-2005,ReinhardSalvatoreReview}. 

In some respects, one-dimensional systems behave qualitatively different from their three-dimensional counterparts. Most importantly, they generically display the perfect-nesting instability
because the two Fermi points at half band-filling
are connected by half a reciprocal lattice vector. Umklapp scattering turns the 
system insulating as soon as the (effective) interaction 
of the particles becomes finite~\cite{Giamarchi}. Therefore, $U_{\rm c}=0^+$ is the generic situation~\cite{LiebWu,Gebhardbook,Essler},
in contrast to Mott's expectations.
Correspondingly, the phase diagram for the one-dimensional Hubbard model 
does not contain a finite metallic region. When the Hubbard model is extended by the
inclusion of a nearest-neighbor interaction, the ground-state phase
diagram becomes more varied but one can only study quantum phase
transitions between Mott-Hubbard, charge-density-wave (CDW) insulator,
and bond-order-wave (BOW) insulator~\cite{Jeckelmann2002,Satoshis2007,MundNoackLegeza2009}. 
For more information on density waves in strongly correlated quantum chains, see
Ref.~\cite{HohenadlerFehske2018}.

To avoid Umklapp scattering at weak coupling, one can investigate models with only 
right-moving electrons that display only one Fermi point.
A known example is the $1/r$-Hubbard model with its linear dispersion relation 
within the Brillouin zone~\cite{GebhardRuckenstein,GebhardGirndtRuckenstein,Gebhardbook}.
Indeed, as indicated analytically~\cite{GebhardRuckenstein,GebhardBares1995} and
recently corroborated using DMRG~\cite{GebhardLegezaOneoverrHM},
the critical interaction strength for the Mott transition is finite
in the $1/r$-Hubbard model. 

Therefore, we can study the competition of the metallic and insulating phases and 
the corresponding quantum phase transitions using the extended $1/r$-Hubbard model
in one dimension. The resulting phase diagram should be generic in the sense
that each phase covers a finite region in the ground-state phase diagram,
as is expected for a three-dimensional system at half band-filling
without Umklapp scattering.

\subsection{Phase diagram}
\label{subsec:phasediagram}

\begin{figure}[ht]
\includegraphics[width=8cm]{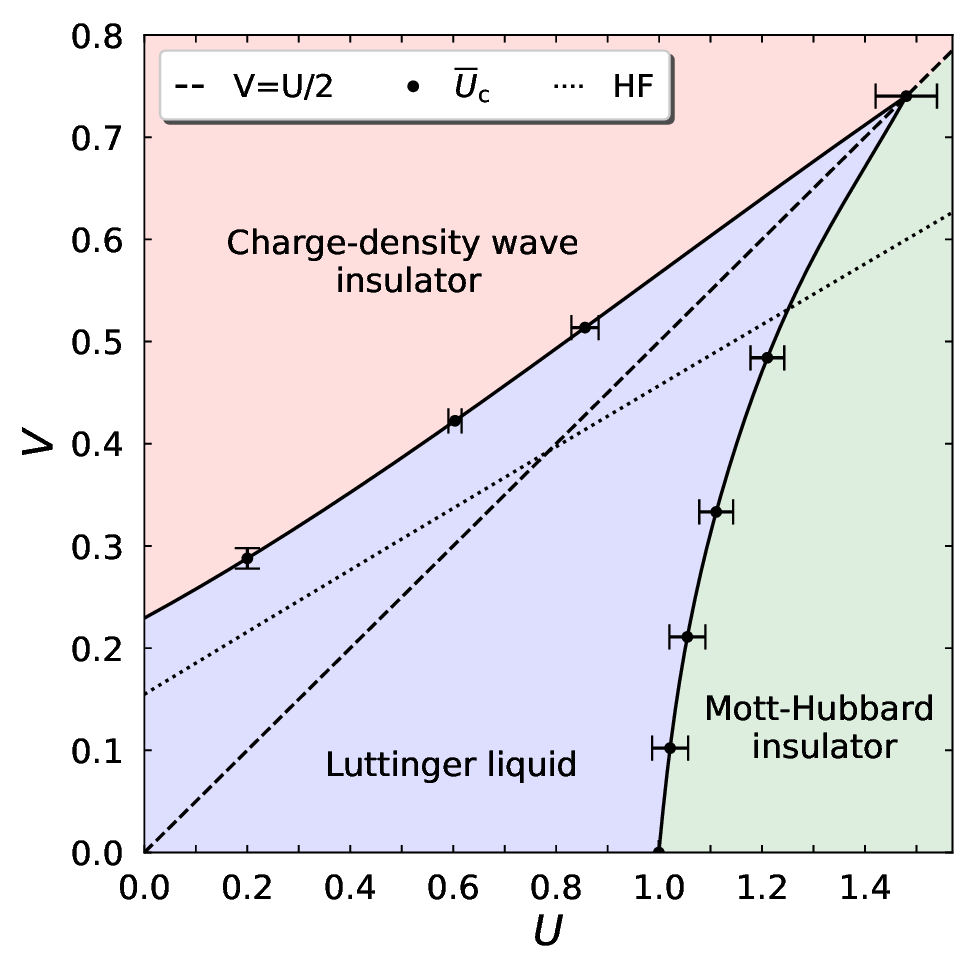}
\caption{Phase diagram of the one-dimensional extended $1/r$-Hubbard model; energies in units of the bandwidth, $W=1$. Dots: estimate for the critical interaction, $\bar{U}_{\rm c}$, with error bounds; continuous lines:
spline interpolations through the dots as guide to the eye; dotted line: Hartree-Fock (HF) result for
the transition between metal and charge-density-wave insulator.\label{fig:phasediagram}}
\end{figure}

The phase diagram in Fig.~\ref{fig:phasediagram} depicts the central result of our work.
It shows the generic Mott-Hubbard phase diagram 
for extended Hubbard models without Umklapp scattering.
Derived for the special case of the extended $1/r$-Hubbard model, 
the phase diagram displays finite regions for
the generic phases of an interacting electron system
with a single half-filled band of width~$W\equiv 1$
and with tunable local interaction~$U$ and nearest-neighbor interaction~$V$.

As can be argued using weak-coupling and strong-coupling perturbation theory,
there should be a metallic phase at weak interactions, $U,V \ll W$, that
becomes unstable against 
a Mott-Hubbard insulator for dominant Hubbard interaction, $U\gg V,W$,
or against a charge-density-wave (CDW) insulator for dominant nearest-neighbor
interactions, $V \gg U,W$. The critical interactions for the corresponding
quantum phase transitions should be finite, the competing interactions being of the same
order of magnitude.

Indeed, when the Coulomb interactions are dominant, $U,V \gg W$,
the separation line between Mott-Hubbard insulator and CDW insulator
should be $V=U/2$. The corresponding curve is included as a dashed line in Fig.~\ref{fig:phasediagram}.
For large $U,V$, we find $V_{\rm c}(U)\gtrsim U/2$,
with small deviations in favor of the Mott-Hubbard insulator. For this reason, we only show the phase diagram for $U\leq 1.6$.
A bond-order wave might separate the two insulating phases,
as is found in the one-dimensional extended Hubbard model~\cite{Jeckelmann2002,Satoshis2007,MundNoackLegeza2009}.
Therefore, the line separating Mott-Hubbard insulator and charge-density-wave insulator 
should be taken as a guide to the eye only.

In our study we focus on the transitions between the metallic Luttinger liquid and the two insulating phases.
We determine $V_{\rm c}(U)$ for fixed $0\leq v=V/U\leq 0.7$ with increment $\Delta v =0.1$, and for fixed $U=0.2$;
for the meaning of the error bars in Fig.~\ref{fig:phasediagram}, 
see Sect.~\ref{sec:MTnearest-neighbor}. 

We note the following.
\begin{itemize}
    \item[--] In the absence of a nearest-neighbor interaction, the Mott-Hubbard  transition 
    is known to occur at 
    $U_{\rm c}(V=0)=1$~\cite{GebhardRuckenstein,Gebhardbook} which is
well reproduced using DMRG~\cite{GebhardLegezaOneoverrHM}. The
repulsive nearest-neighbor interaction {\em increases\/} the critical interaction strength, 
i.e., the inclusion of the nearest-neighbor interaction stabilizes the {\em metallic\/} phase. 
Apparently, the additional repulsive nearest-neighbor interaction 
softens the two-particle scattering potential that is purely local in the bare Hubbard model.

As a major result we find that the Mott transition remains continuous in the presence of a nearest-neighbor
interaction. We presume that short-range interactions that decrease as a function of the particle distance will not 
fundamentally alter this behavior. 
\item[--] The transition from the Luttinger liquid metal to the charge-density-wave insulator is fairly common in the sense that 
even Hartree-Fock theory qualitatively reproduces the transition for not too large interactions. 

In Fig.~\ref{fig:phasediagram}, the corresponding Hartree-Fock prediction
is shown as a dotted line.
As usual, Hartree-Fock theory overestimates the stability of the ordered state and thus underestimates 
the critical interaction, $V_{\rm c,CDW}^{\rm HF}(U)<V_{\rm c,CDW}(U)$.
Since the metallic phase extends well beyond the line $V=U/2$, there is no indication 
for a bond-order wave that might separate the Luttinger liquid and the charge-density-wave insulator.
\end{itemize}
We use a third-order spline interpolation through the data points to draw 
the phase transition lines in Fig.~\ref{fig:phasediagram}. 
The full lines depict continuous quantum phase transitions in the sense that 
the gaps open and close continuously at the same critical interaction
when the transition is approached from the metallic and insulating sides, respectively.

The endpoint of both continuous lines where all three phases meet deserves spatial attention.
Unsurprisingly, finite-size corrections are most severe in this region of phase space, 
and the study of the region around the tri-critical point is cumbersome and beyond the scope of
our presentation. 

\subsection{Outline}
\label{subsec:outline}

Our work is organized as follows. In Sect.~\ref{sec:modeldef} we define the
Hubbard model with long-range electron transfers and
onsite and nearest-neighbor Coulomb interactions.
We introduce the ground-state properties of interest, namely,
the ground-state energy, the two-particle gap, the
momentum distribution, and the density-density correlation function
from which we determine the Luttinger parameter in the metallic phase
and the CDW order parameter. 
In Sect.~\ref{sec:gsproperties} we present results for the ground-state properties
and discuss their finite-size dependencies and extrapolations to the thermodynamic limit where appropriate. 

In Sect.~\ref{sec:MTnearest-neighbor}
we focus on the Mott-Hubbard transition in the presence of a nearest-neighbor
interaction. We propose and discuss several methods to extract the critical interaction
strength for the Mott transition based on the ground-state energy, the two-particle gap, the Luttinger parameter, and the 
structure factor whereby we study the Mott transition at fixed
$v\equiv V/U$ in the range $0\leq v \leq 0.7$ (increment $\Delta v=0.1$) in units of the bandwidth, $W\equiv 1$.
In addition, we address the Mott transition as a function of $V$ for fixed $U=0.2$ and $U=1.7$.

Short conclusions, Sect.~\ref{sec:conclusions}, close our presentation. The Hartree-Fock calculations 
for the CDW transition are collected in the appendix.

\section{Hubbard model with linear dispersion}
\label{sec:modeldef}

\subsection{Hamiltonian}

In this work, we address the $1/r$-Hubbard model~\cite{GebhardRuckenstein,Gebhardbook}
with nearest-neighbor interactions
\begin{equation}
  \hat{H}=\hat{T}+U\hat{D}+V \hat{V} 
  \label{eq:fullHubbardmodel}
\end{equation}
on a ring with $L$~sites ($L$: even).
We discuss the kinetic energy and the Coulomb interaction terms separately.

\subsubsection{Kinetic energy}

The kinetic energy describes the tunneling of electrons with
spin~$\sigma=\uparrow,\downarrow$ along a ring with $L$ sites,
\begin{eqnarray}
      \hat{T} &=& \sum_{\substack{l,m=1\\
      l\neq m; \sigma}}^{L}t(l-m)
    \hat{c}_{l,\sigma}^+\hat{c}_{m,\sigma}^{\vphantom{+}} \; ,
      \label{eq:defT} \\
  t(r) &=& (-\rmi t ) \frac{(-1)^r}{d(r)} \; ,\nonumber \\
  d(r)& =& \frac{L}{\pi}\sin\left(\frac{\pi r}{L}\right) \; .
  \label{eq:defTconstituents}
\end{eqnarray}
The creation and annihilation operators $\hat{c}_{l,\sigma}^{+}$,
  $\hat{c}_{l,\sigma}^{\vphantom{+}}$
  for an electron with spin
$\sigma=\uparrow,\downarrow$ on lattice site~$l$ obey the usual
anti-commutation relations for fermions.

In Eq.~(\ref{eq:defTconstituents}),
$d(l-m)$ is the chord distance between the sites~$l$ and $m$ on a ring.
In the thermodynamic limit and for $|l-m|\ll L $ fixed,
we have $d(l-m)= (l-m)+ {\cal O}(1/L^2)$, and the electron
transfer amplitude between two sites decays inversely proportional
to their distance (`$1/r$-Hubbard model').

Since $L$ is even,
we have anti-periodic electron transfer amplitudes
because $d(L+ r)= -d(r)$. Therefore, we must choose
anti-periodic boundary conditions
\begin{equation}
\hat{c}_{L+l,\sigma}=-\hat{c}_{l,\sigma}
\end{equation}
for the operators, too.
With these boundary conditions,
the kinetic energy operator is diagonal in Fourier space,
\begin{eqnarray}
  \hat{C}_{k,\sigma}^+
  &= & \frac{1}{\sqrt{L}}\sum_{l= 1}^{L}
  e^{\rmi  kl } \hat{c}_{l,\sigma}^+ \; , \nonumber \\
  \hat{c}_{l,\sigma}^{+}
    &= & \frac{1}{\sqrt{L}}\sum_k
    e^{-\rmi  kl } \hat{C}_{k,\sigma}^+ \; ,    \nonumber \\
    k&=& \frac{(2m+ 1)\pi}{L}\;, \; m= -\frac{L}{2}, \ldots, \frac{L}{2}-1\; ,
    \label{eq:FTofoperators}
\end{eqnarray}
so that
\begin{equation}
  \hat{T}=\sum_{k,\sigma}\epsilon(k) 
  \hat{C}_{k,\sigma}^+\hat{C}_{k,\sigma}^{\vphantom{+}}\; , \quad
\epsilon(k)=t k \; .
\end{equation}
The dispersion relation of the $1/r$-Hubbard model is linear. We set
\begin{equation}
t=\frac{1}{2\pi}
\end{equation}
so that the bandwidth is unity, $W\equiv 1$.

In  this work, we focus on the case of a paramagnetic half-filled ground state
where we have the same number of electrons per spin species, 
$N_{\uparrow}= N_{\downarrow}$,
that equals half the number of lattice sites, $N_{\sigma}=L/2$
($\sigma=\uparrow,\downarrow$).

\subsubsection{Coulomb interaction}

The Coulomb interaction is parameterized by two terms
in Eq.~(\ref{eq:fullHubbardmodel}).
The on-site (Hubbard) interaction~\cite{Hubbard1963,Gutzwiller1963,Kanamori}
acts locally between two electrons with opposite spins,
\begin{equation}
  \hat{D}= \sum_{l=1}^L \hat{n}_{l,\uparrow}\hat{n}_{l,\downarrow}
  \; ,  \quad
\hat{n}_{l,\sigma}=\hat{c}_{l,\sigma}^+\hat{c}_{l,\sigma}^{\vphantom{+}}
  \; ,
\end{equation}
where $\hat{n}_{l,\sigma}$
counts the number of electrons with spin $\sigma$ on site~$l$,
and $\hat{n}_l=\hat{n}_{l,\uparrow}+\hat{n}_{l,\downarrow}$
counts the number of electrons
on site~$l$. The corresponding operators for the total number
of electrons with spin~$\sigma=\uparrow,\downarrow$
are denoted by $\hat{N}_{\sigma}=\sum_l\hat{n}_{l,\sigma}$,
and $\hat{N}=\hat{N}_{\uparrow}+\hat{N}_{\downarrow}$.

To discuss the influence of the extended nature of the Coulomb interaction,
we consider the case of pure nearest-neighbor interactions,
\begin{equation}
  \hat{V}=
  \sum_{l=1}^{L} (\hat{n}_l-1)(\hat{n}_{l+1}-1)\; ,
  \label{eq:NNinteraction}
\end{equation}
where we disregard the long-range parts of the Coulomb interaction
for distances $|l-m|\geq 2$. 
The model in Eq.~(\ref{eq:fullHubbardmodel})
describes the `extended' $1/r$-Hubbard model with on-site interaction~$U$
and nearest-neighbor interaction~$V$.

As we shall show in this work, 
the Mott-Hubbard transition at half band-filling remains continuous
in the presence of short-range interactions. For not too large interactions and for $V\lesssim U/2$,
the model contains a transition from the Luttinger-liquid metal  to the Mott-Hubbard insulator.
For larger nearest-neighbor interactions, the model eventually describes
transitions from the metallic state to 
a charge-density-wave (CDW) insulator.
For strong interactions, $U\gg W$, the model contains a transition from the Mott-Hubbard insulator
to the CDW insulator around $V\approx U/2$.

We study several values for the ratio $v= V/U$,
namely, $v=0, 0.1,0.3,0.4,0.5,0.6,0.7$ for weak to strong nearest-neighbor interactions.
Since we scan the value of $U$,
we must limit the number of values for $v$ to keep the numerical effort within bounds when we include systems up to $L_{\rm max}=80$ lattice sites; when finite-size effects are well behaved,
 e.g., for the ground-state energy,
we limit our investigations to $L=64$.
Moreover, we scan $V$ for fixed $U=0.2$ and $U=1.7$ to study the Mott transition as a function of
the nearest-neighbor interaction.

\subsubsection{Particle-hole symmetry}

Under the particle-hole transformation
\begin{equation}
  \hat{c}_{l,\sigma}^{\vphantom{+}} \mapsto \hat{c}_{l,\sigma}^+ \quad
  , \quad \hat{n}_{l,\sigma} \mapsto 1-\hat{n}_{l,\sigma} \; ,
\end{equation}
the kinetic energy remains unchanged,
\begin{eqnarray}
\hat{T} &\mapsto& \sum_{\substack{l,m=1\\      l\neq m; \sigma}}^{L}
t(l-m)\hat{c}_{l,\sigma}^{\vphantom{+}}\hat{c}_{m,\sigma}^+
\nonumber \\
&=&
 \sum_{\substack{l,m=1\\      l\neq m; \sigma}}^{L}
  \left[ -t(m-l)\right]
  \hat{c}_{l,\sigma}^+\hat{c}_{m,\sigma}^{\vphantom{+}}
  = \hat{T}
\end{eqnarray}
  because $t(-r)=-t(r)$.
Furthermore,
\begin{equation}
  \hat{D}\mapsto \sum_{l=1}^L
  (1-\hat{n}_{l,\uparrow})(1-\hat{n}_{l,\downarrow})
  =\hat{D}-\hat{N}+L \; ,
\end{equation}
and
\begin{equation}
  \hat{V}\mapsto \hat{V} \;. 
\end{equation}
Therefore, $\hat{H}(N_{\uparrow},N_{\downarrow})$ has the same spectrum as
$\hat{H}(L-N_{\uparrow},L-N_{\downarrow})-U(2L-N)+LU$,
where $N=N_{\uparrow}+N_{\downarrow}$ is the particle number.

\subsection{Ground-state properties}
\label{subsec:gsprop}

We are interested in the metal-insulator transition at half band-filling where
the metallic Luttinger liquid for weak interactions turns
into a paramagnetic Mott insulator
for large interactions at some finite value $U_{\rm c}(V)$ when $V$ is small enough, 
or to a CDW insulator for strong nearest-neighbor interactions.
The metal-insulator transition can be inferred from the finite-size extrapolation of the ground-state energy and of
the two-particle gap~\cite{GebhardLegezaOneoverrHM}. Alternatively,
the Luttinger parameter~\cite{PhysRevB.106.205133} and 
the finite-size extrapolation of the structure factor at the Brillouin zone boundary permit to determine
the critical interaction strength.
Moreover, the charge-density-wave state can be monitored by the CDW order parameter.
In this section, we also introduce the momentum distribution
for finite systems that is also accessible via DMRG.

\subsubsection{Ground-state energy and two-particle gap}

We denote the ground-state energy by
\begin{equation}
E_0(N,L;U,V)= \langle \Psi_0 | \hat{H} |\Psi_0 \rangle
\end{equation}
for given particle number~$N$, system size~$L$, and interaction
parameters $U,V$.
Here, $|\Psi_0\rangle$ is the normalized ground state of the
Hamiltonian~(\ref{eq:fullHubbardmodel}).
We are interested in the thermodynamic limit,
$N, L\to \infty$ with $n=N/L$ fixed.
We denote the ground-state energy
per site and its extrapolated value by
\begin{eqnarray}
  e_0(N,L;U,V)&=& \frac{1}{L} E_0(N,L;U,V) \;, \nonumber \\
  e_0(n;U,V)&= & \lim_{L\to \infty} e_0(N,L;U,V) \; ,
\end{eqnarray}
respectively.

The two-particle gap is defined by
\begin{equation}
  \Delta_2(L;U,V) = \mu_2^+(L;U,V)-\mu_2^-(L;U,V) \; ,
  \label{eq:tpgapdef}
\end{equation}
where
\begin{eqnarray}
  \mu_2^-(L;U,V)&=& E_0(L,L;U,V)- E_0(L-2,L;U,V)
  \nonumber \; , \\
  \mu_2^+(L;U,V)  &=& E_0(L+ 2,L;U,V)- E_0(L,L;U,V)
    \nonumber \\
  \label{eq:defmu2plus}
\end{eqnarray}
are the chemical potentials for adding the last two particles to half filling
and the first two particles beyond half filling, respectively.

Due to particle-hole symmetry, we have
\begin{equation}
  \mu_2^-(L;U,V)= 2U-\mu_2^+(L;U,V)
\end{equation}
so that
\begin{equation}
  \Delta_2(L;U,V) = 2\mu_2^+(L;U,V)-2U
  \label{eq:defDelta2}
\end{equation}
and
\begin{equation}
  \Delta_2(U,V) = \lim_{L\to\infty} \Delta_2(L;U,V)
  \label{eq:defDeltaTDlim2}
\end{equation}
in the thermodynamic limit.
We always consider the spin symmetry sector $S=S^z=0$. For this reason,
we study the two-particle gap rather than the single-particle gap.

The two added particles repel each other so that, in the thermodynamic limit,
they are infinitely separated from each other. Therefore, we have
\begin{equation}
\Delta_2(U,V)=2 \Delta_1 (U,V) \; ,
\end{equation}
where $\Delta_1(U,V)$ is the gap for single-particle excitations.
For finite systems, we expect the interaction energy
\begin{equation}
  e_{\rm R}(L;U,V) = \Delta_2(L;U,V)-2 \Delta_1 (L;U,V)={\cal O}(1/L)>0
  \label{eq:defdeltaeR}
\end{equation}
to be positive, of the order $1/L$.
We verified that the interaction energy vanishes in the thermodynamic limit
for the case $V=0$~\cite{GebhardLegezaOneoverrHM}.
  
\subsubsection{Momentum distribution}

We also study the spin-summed momentum distribution in the ground state
at half band-filling, $N=L$,
\begin{eqnarray}
  n_k(L;U,V) &=& \langle \Psi_0 | \hat{n}_{k,\uparrow} + \hat{n}_{k,\downarrow}
  |\Psi_0\rangle \nonumber \\
  &= & \sum_{l,m;\sigma} e^{\rmi k (l-m)} P_{l,m;\sigma}
  \label{eq:defmomentumdistribution}
\end{eqnarray}
with $\hat{n}_{k,\sigma}=\hat{C}_{k,\sigma}^+\hat{C}_{k,\sigma}^{\vphantom{+}}$
and the single-particle density matrix $P_{l,m;\sigma}=
\langle \Psi_0 |\hat{c}_{l,\sigma}^+ \hat{c}_{m,\sigma}^{\vphantom{+ }} |\Psi_0\rangle$.
Due to particle-hole symmetry we have
\begin{equation}
n_k(L;U,V)= 1-n_{-k}(L;U,V)
\end{equation}
at half band-filling. Therefore, it is sufficient to study wave numbers from
the interval $-\pi<k<0$.

In contrast to our previous work~\cite{GebhardLegezaOneoverrHM},
the slope of the momentum distribution at the band edge cannot be used
to trace the Mott-Hubbard transition in the extended
$1/r$-Hubbard model because the bound state moves away from the band edge
for $V>0$.

\subsubsection{Density-density correlation function and Luttinger parameter}
\label{subsec:CNNexact}

Lastly, we address the density-density correlation function
at half band-filling, $N= L$,
\begin{equation}
  C^{\rm NN}(r,L;U,V) = \frac{1}{L} \sum_{l=1}^L
  \bigl(\langle\hat{n}_{l+r}\hat{n}_l \rangle
  - \langle \hat{n}_{l+r}\rangle \langle \hat{n}_l \rangle\bigr) \; ,
  \label{eq:CNNdef}
\end{equation}
where $\langle \ldots \rangle \equiv \langle \Psi_0| \ldots | \Psi_0\rangle$.
The limit $L\gg r\gg 1$ for $U,V \ll W$
is also accessible
from field theory~\cite{Thierrybook,PhysRevLett.64.2831,PhysRevB.39.4620},
\begin{equation}
  C^{\rm NN}(r\gg 1;U,V) \sim - \frac{K(U,V)}{(\pi r)^2}
  +\frac{A(U,V) (-1)^r}{r^{1+K}[\ln(r)]^{3/2}} + \ldots \, ,
  \label{eq:CNNfieldtheory}
\end{equation}
where $A(U,V)$ is a constant that depends on the interaction but not
on the distance~$r$.

We extract the Luttinger exponent~$K(U,V)$ from the structure factor,
\begin{equation}
  \tilde{C}^{\rm NN}(q,L;U,V)
  = \sum_{r=0}^{L-1}e^{-{\rm i}q r} C^{\rm NN}(r,L;U,V) 
  \;,
  \label{eq:CNNtildedef}
\end{equation}
where the wave numbers are from momentum space,
$q=(2\pi/L)m_q$, $m_q=-L/2,-L/2+1,\ldots,L/2-1$.
By construction, $\tilde{C}^{\rm NN}(q=0,L;U,V)=0$
because the particle number is fixed,
$N=L$ in the half-filled ground state. 
In the thermodynamic limit, the structure factor $\tilde{C}^{\rm NN}(q,L;U,V)$ remains of the order unity even in the CDW phase because we subtract
the contributions of the long-range order in the definition~(\ref{eq:CNNdef}).

The transition to a charge-density-wave insulator can be monitored from the CDW order parameter.
In this work, we do not study the standard CDW order parameter,
\begin{equation}
    D(L;U,V)=  \frac{1}{L} \left| \sum_{r=0}^{L-1} (-1)^r \left(\langle \hat{n}_{r} \rangle -1 \right)\right|\leq 1\; .
    \label{eq:CDWorderparameterdef}
\end{equation}
Instead, we include all short-range contributions and address
\begin{equation}
    N_{\pi}(L;U,V)=  \frac{1}{L} \sum_{r=0}^{L-1} (-1)^r \frac{1}{L} \sum_{l=0}^{L-1} \left(\langle \hat{n}_{r+l} \hat{n}_{l}\rangle -1 \right) \; .
    \label{eq:ourCDWorderparameterdef}
\end{equation}
When the charges are distributed homogeneously, $\langle \hat{n}_l\rangle=1$, we have
$N_{\pi}(L;U,V)=\tilde{C}^{\rm NN}(\pi,L;U,V)/L$, and the order parameter vanishes in the metallic phase.
More generally, in the thermodynamic limit we have $N_{\pi}(U,V)=(D(U,V))^2$. In the $1/r$-Hubbard model
with its long-range electron transfer, it is advantageous to analyze $N_{\pi}(L;U,V)$
to facilitate a reliable finite-size analysis. 

When Eq.~(\ref{eq:CNNfieldtheory}) is employed, it follows
that the Luttinger parameter for finite systems,
\begin{equation}
  K(L;U,V)
  = \frac{L}{2} \tilde{C}^{\rm NN}(2\pi/L,L;U,V) \; ,
  \label{eq:KfromCNNfinite}
\end{equation}
can be used to calculate the Luttinger parameter in the thermodynamic limit,
\begin{eqnarray}
  K(U,V)&=& \lim_{L\to\infty} K(L;U,V)\nonumber \\
  &=& \pi \lim_{q\to 0}\frac{\tilde{C}^{\rm NN}(q;U,V)}{q}\; ,
  \label{eq:KfromCNN}
\end{eqnarray}
where we denote the structure factor
in the thermodynamic limit by $\tilde{C}^{\rm NN}(q;U,V)$.
Using Eq.~(\ref{eq:KfromCNN}), the Luttinger exponent can be
calculated numerically
with very good accuracy~\cite{Ejima2005}.
The Luttinger parameter can be used to locate the metal-insulator transition
in one spatial dimension. 

\section{Ground-state properties}
\label{sec:gsproperties}

Before we investigate the Mott transition for the half-filled extended $1/r$ Hubbard model
in more detail in the next section,
we present DMRG results for the ground-state energy, the two-particle gap, the momentum distribution,
the structure factor, and the CDW order parameter. 
For the numerical calculations we employ a DMRG code that permits the treatment of arbitrary quantum system with long-ranged complex interactions. It uses non-Abelian symmetries and optimization protocols inherited from quantum information theory~\cite{Szalay-2015b}.
Further technical details of the DMRG implementation can be
found in Ref.~\cite{GebhardLegezaOneoverrHM}.
Note that our finite-size scaling analysis requires very accurate data. We obtain those
by imposing strict accuracy settings in our DMRG code, and by restricting 
the largest system size to $L_{\rm max}=80$ to limit the truncation errors.

\subsection{Ground-state energy}

For $V= 0$, the ground-state energy per site for finite system sizes 
is given by ($n=N/L$,
$N$: even)~\cite{GebhardRuckenstein,Gebhardbook,GebhardLegezaOneoverrHM}
\begin{eqnarray}
  e_0 &=&\frac{1}{4} n(n-1) +\frac{U}{4} n   \label{eq:gsenergyNeven} \\
  && -\frac{1}{2L} \sum_{r=0}^{(N/2)-1}\sqrt{1+U^2-4U(2r+1-L/2)/L} \nonumber 
\end{eqnarray}
with the abbreviation $e_0\equiv e_0(N,L;U,V= 0)$.

In  the thermodynamic limit and at half band-filling, $n= 1$,
the ground-state energy per site becomes particularly simple,
\begin{eqnarray}
  e_0(n= 1;U\leq 1,V= 0)
  &=& -\frac{1}{4} + \frac{U}{4}-\frac{U^2}{12} \; ,\nonumber \\
  e_0(n= 1;U\geq 1,V= 0) &=& -\frac{1}{12 U} \; .
  \label{eq:e0V0TDLhalffilling}
\end{eqnarray}
The analytic expressions~(\ref{eq:gsenergyNeven})
and~(\ref{eq:e0V0TDLhalffilling}) are useful for a
comparison with numerical data at $V=0$.

For finite~$V$, we can use first-order perturbation theory for weak interactions, $U,V\ll1$, to find
\begin{figure}[t]
 \begin{flushleft}
(a)  \includegraphics[width=7cm]{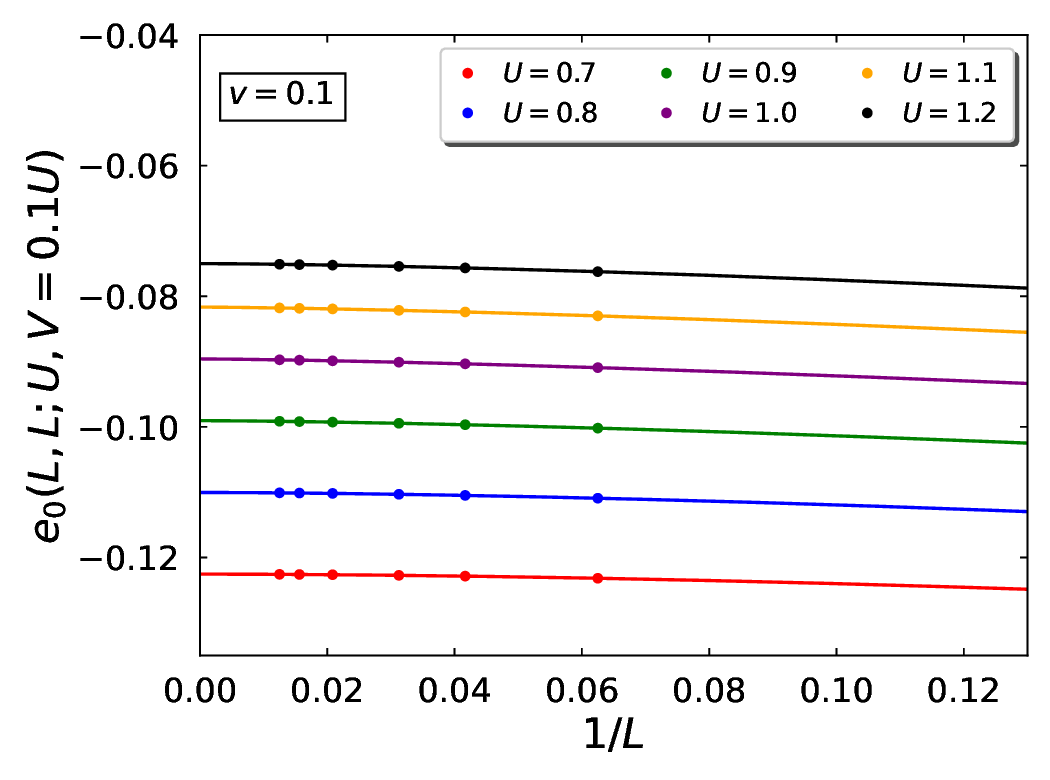}\\[12pt]
(b)  \includegraphics[width=7cm]{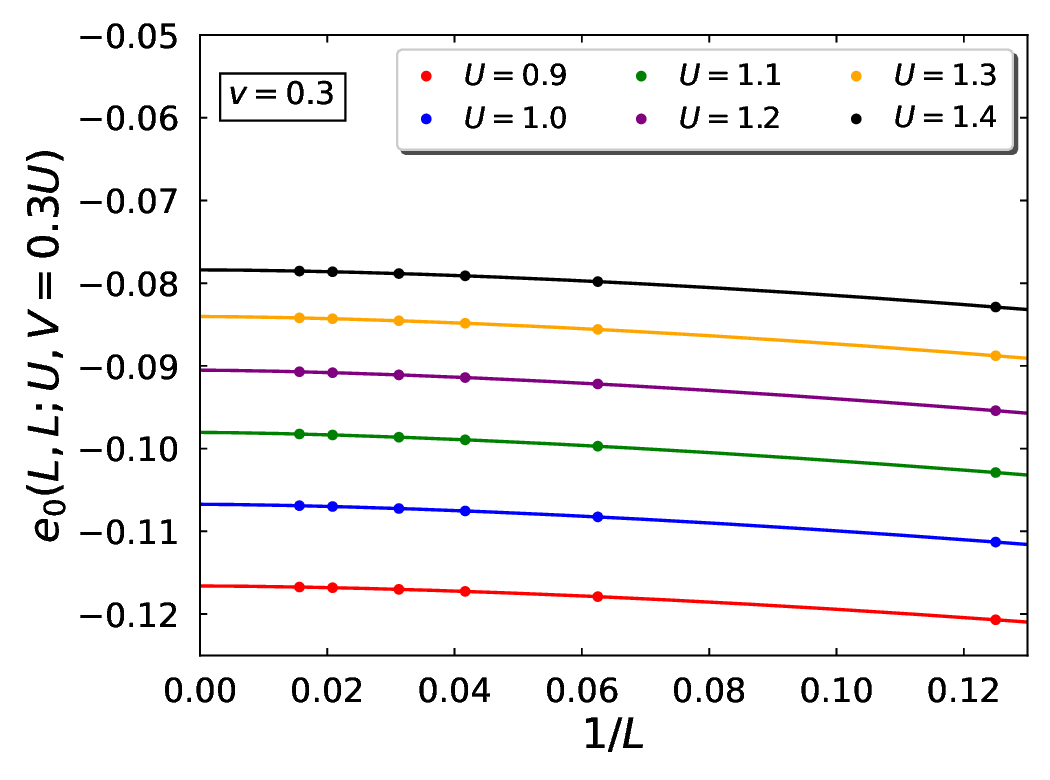}\\[12pt]
(c)  \includegraphics[width=7cm]{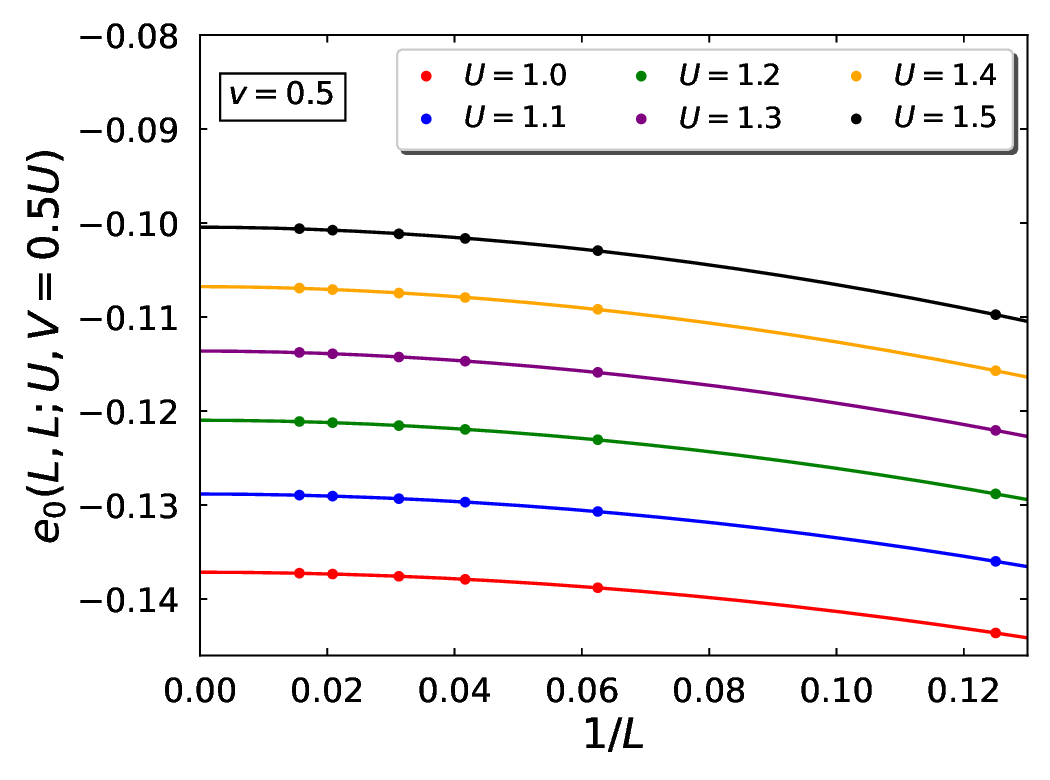}
 \end{flushleft}
 \caption{Ground-state energy per lattice site
   at half band-filling, $e_0(L,L;U,V)$,
   for the extended $1/r$-Hubbard model
   as a function of $1/L$ for $L=8,16,24,32,48,64$
   and various values for $U$
    for (a) $v=0.1$, (b) $v=0.3$, (c) $v=0.5$.
    The continuous lines are fits to the algebraic
    fit function~(\protect\ref{eq:ezeroextrapolation}).
The intercept of the extrapolation curves with the ordinate
    defines the extrapolation estimate $e_0(n=1;U,V)$
in the thermodynamic limit.\label{fig:gsenergy}}
\end{figure}
\begin{equation}
    e_0^{\rm PT}(U,V)=-\frac{1}{4}+\frac{U}{4}\left(1-\frac{8v}{\pi^2}\right)+\mathcal{O}(U^2)
    \label{eq:e01storderpt}
\end{equation}
with $v=V/U$ in the thermodynamic limit and at half band-filling. Note that Eq.~(\ref{eq:e01storderpt}) holds for all $v$.

We display the ground-state energy per site
at half band-filling, $e_0(L,L;U,V)$, 
as a function of the inverse system size ($L= 8,16,24,32,48,64$)
and various values of $U$ in Fig.~\ref{fig:gsenergy}a ($v=0.1$),
Fig.~\ref{fig:gsenergy}b ($v=0.3$), and Fig.~\ref{fig:gsenergy}c ($v=0.5$).
For the extrapolation to the thermodynamic limit, we use the
algebraic fit function
\begin{equation}
  e_0(L,L;U,V) = e_0(n=1;U,V) + a_0(U,V) \left(\frac{1}{L}\right)^{\gamma_0(U,V)} \; ,
  \label{eq:ezeroextrapolation}
  \end{equation}
where $e_0(n=1;U,V)$ denotes the 
numerical estimate for the ground-state energy density
in the thermodynamic limit and $a_0(U,V)$ and $\gamma_0(U,V)$ are the two other
fit parameters.
This extrapolation scheme is appropriate
for $V= 0$~\cite{GebhardLegezaOneoverrHM} because the ground-state energy
per site scales with $(1/L)^2$ for $U\neq 1$ and with $(1/L)^{3/2}$
for $U= U_{\rm c}(V= 0)= 1$, as follows from Eq.~(\ref{eq:gsenergyNeven}).
More generally, we assume for all $(U,V)$
\begin{equation}
    \gamma_0(U,V) = \left\{ \begin{array}{ccl}
    2 & \hbox{for} & U \neq U_{\rm c}(V) \\[6pt]
    \displaystyle \frac{3}{2} & \hbox{for} & U = U_{\rm c}(V) 
    \end{array}
    \right. \; .
    \label{eq:gamma0TDL}
\end{equation}
These exponents apply for very large system sizes. 
We shall discuss the finite-size modifications
in detail in Sect.~\ref{sec:MTnearest-neighbor}.

\begin{figure}[t]
    \includegraphics[width=8cm]{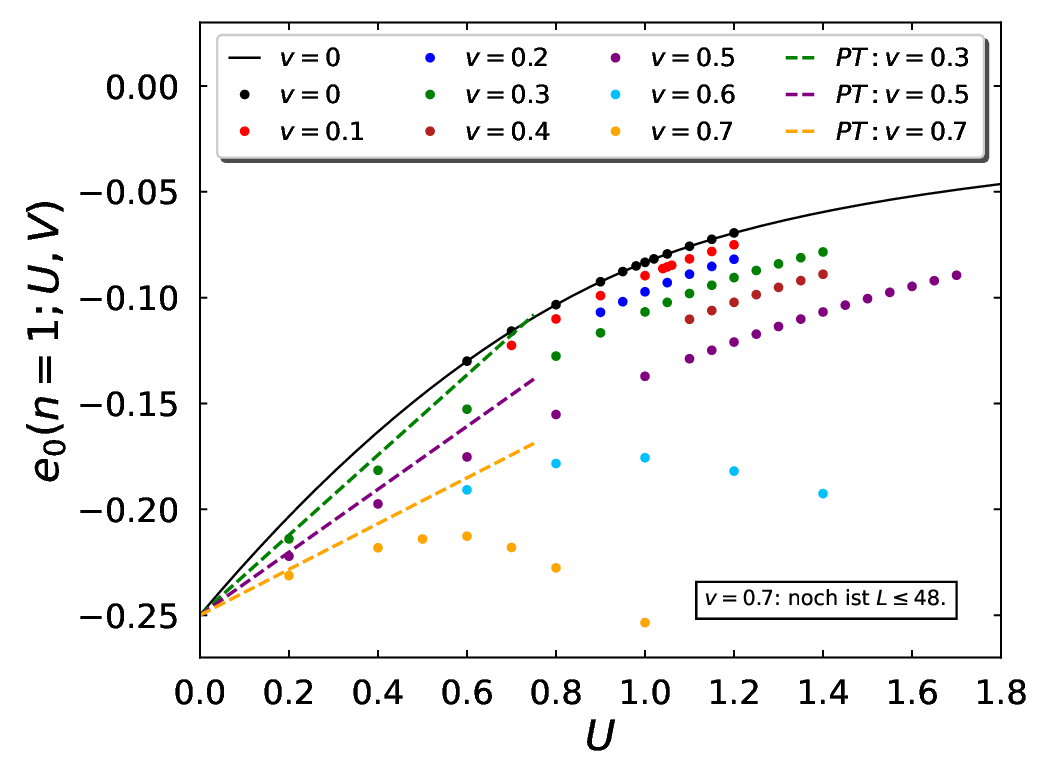}
 \caption{Ground-state energy per lattice site
   at half band-filling in the thermodynamic limit, $e_0(n=1;U,V)$,
   for the extended $1/r$-Hubbard model from the extrapolation
   to the thermodynamic limit in Fig.~\protect\ref{fig:gsenergy}.
   The dashed lines represent first-order order perturbation theory for $v=V/U=0.3,0.5,0.7$, see Eq.~(\ref{eq:e01storderpt}).
   The continuous line is the exact result
   for $V= 0$, $e_0(n=1;U,V=0)$, see
   Eq.~(\protect\ref{eq:e0V0TDLhalffilling}).\label{fig:gsenergyextrapolate}}
\end{figure}

The extrapolated ground-state energies are shown in Fig.~\ref{fig:gsenergyextrapolate}
together with the exact result for $V= 0$.
For small interactions,
the nearest-neighbor interaction in the particle-hole symmetric form
decreases the ground-state energy because
the Hartree contribution at half band-filling is subtracted
in the definition of the interaction, and the Fock contribution
is negative because of the exchange hole.
Therefore, the linear term in the interaction $(U/4)(1-8v/\pi^2)$, see Eq.~(\ref{eq:e01storderpt}),
is smaller in the presence of a nearest-neighbor interaction.

At large interactions,
the ground-state energy approaches zero,  $\lim_{U\to \infty} e_0(n=1;U,V= vU)= 0$, as long as the charge-density wave is absent.
In the presence of a CDW, the ground-state energy
is negative and proportional to $U$, $e_0(U\gg 1,V)=U(1/2-v)$.

\subsection{Two-particle gap}

\begin{figure}[t]
 \begin{center}
(a)  \includegraphics[width=7cm]{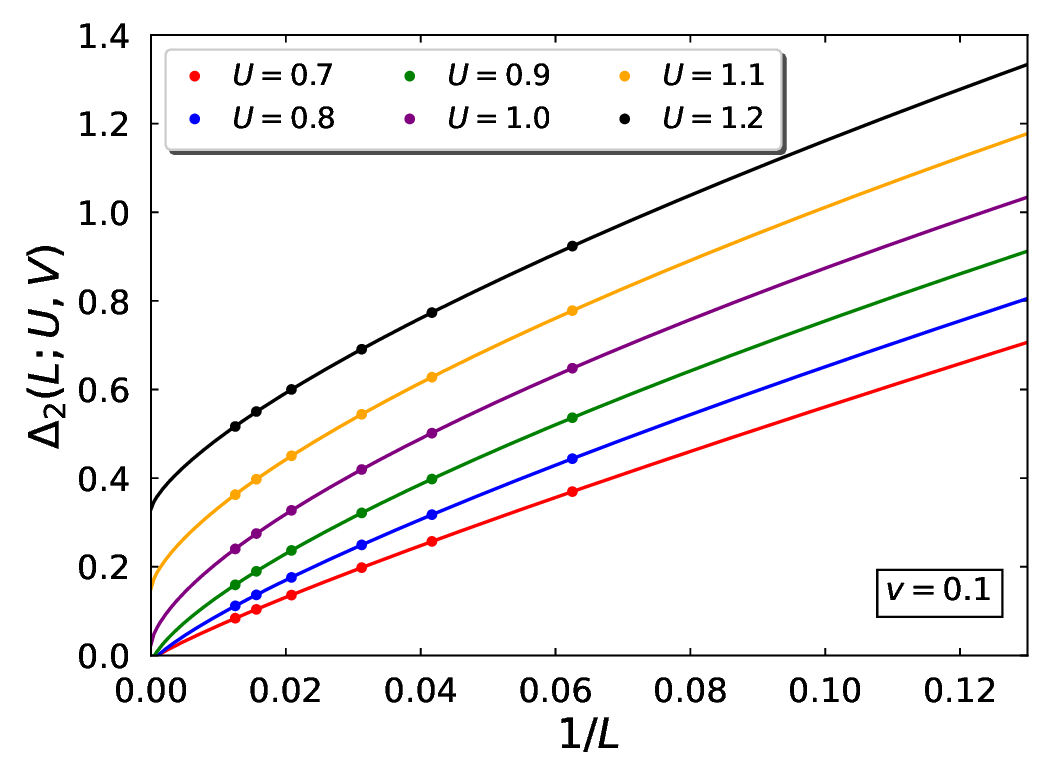}\\[12pt]
(b)  \includegraphics[width=7cm]{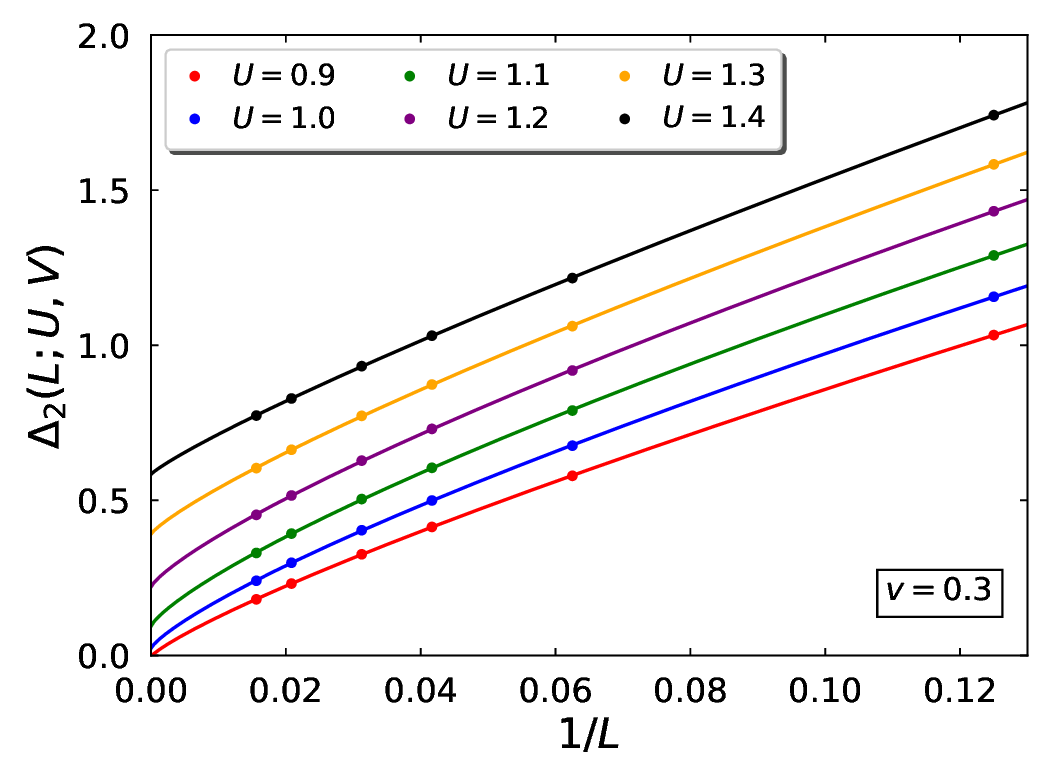}\\[12pt]
(c)  \includegraphics[width=7cm]{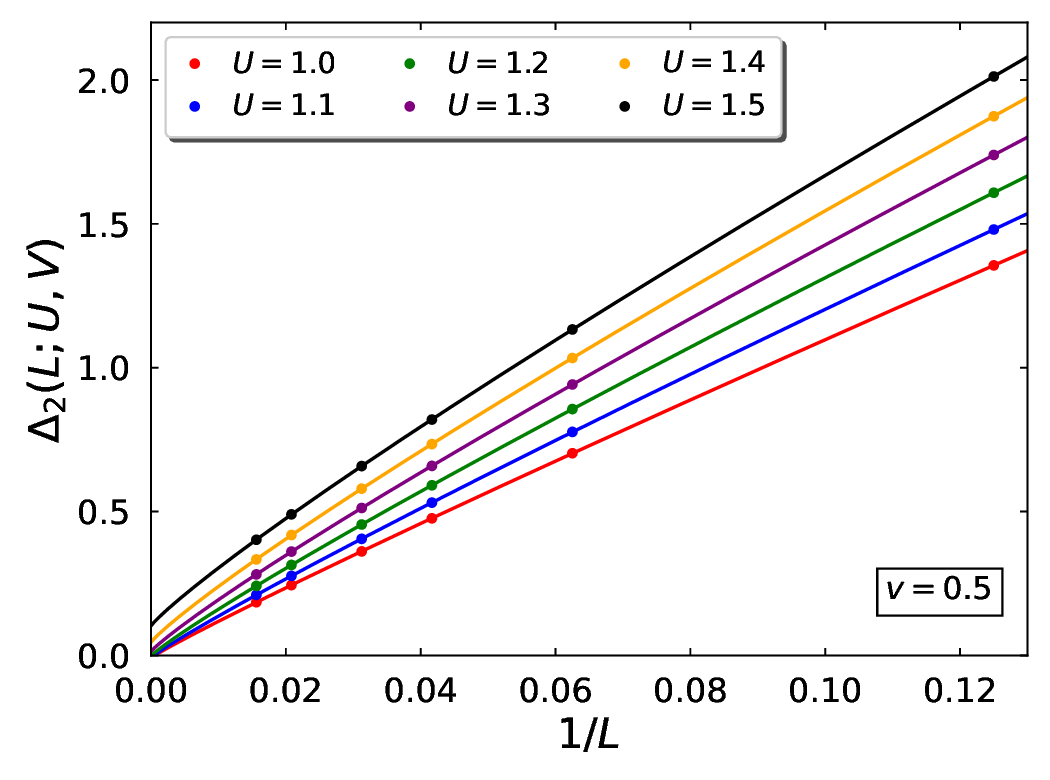}
 \end{center}
 \caption{Two-particle gap $\Delta_2(L;U,V)$
   for the extended $1/r$-Hubbard model
    as a function of inverse system size for $L=8,16,24,32,48,64$ and various values for $U$
    for (a) $v=0.1$, (b) $v=0.3$, (c) $v=0.5$.
    The continuous lines are fits to the algebraic
    fit function~(\protect\ref{eq:gapextrapolationscheme}).
The intercept of the extrapolation curves with the ordinate
    defines the extrapolation estimate $\Delta_2(U,V)$
    for the two-particle gap.\label{fig:Delta2finite}}
\end{figure}

For $V=0$ the two-particle gap is known exactly for all system sizes~\cite{GebhardRuckenstein,Gebhardbook,GebhardLegezaOneoverrHM},
\begin{equation}
    \Delta_2(L;U\geq 1,V=0)=  U-1+\frac{2}{L} +\sqrt{(U-1)^2+\frac{4U}{L}}\; .
    \label{eq:Delta2analyt}
\end{equation}
In the thermodynamic limit, we find
\begin{equation}
    \Delta_2(U\geq 1,V=0) = 2(U-1) \; .
    \label{eq:Delta2TDLVzero}
\end{equation}
The gap opens linearly above the critical interaction strength, $U_{\rm c}(U,V=0)=1$.
Eq.~(\ref{eq:Delta2analyt}) shows that the finite-size data approach the value
in the thermodynamic limit algebraically in $1/L$,
\begin{equation}
  \Delta_2(L;U,V) =\Delta_2(U,V) + a_2(U,V) \left(\frac{1}{L}\right)^{\gamma_2(U,V)}
  \label{eq:gapextrapolationscheme}
  \end{equation}
with $\gamma_2(U\neq U_{\rm c},V=0)=1$ and $\gamma_2(U=U_{\rm c},V=0)=1/2$.

More generally, we assume for all $(U,V)$
\begin{equation}
    \gamma_2(U,V) = \left\{ \begin{array}{ccl}
    1 & \hbox{for} & U \neq U_{\rm c}(V) \\[6pt]
    \displaystyle \frac{1}{2} & \hbox{for} & U = U_{\rm c}(V) 
    \end{array}
    \right. \; .
    \label{eq:gamma2TDL}
\end{equation}
As for the ground-state energy, these exponents apply for very large system sizes. 
We shall discuss the finite-size modifications
in more detail in Sect.~\ref{sec:MTnearest-neighbor}.

In Fig.~\ref{fig:Delta2finite}
we show the DMRG results for $\Delta_2(L;U,V)$ as a function of $1/L$ for
various values for $U$ as a function of $1/L$ for $L=8,16,24,32,48,64$ 
for (a) $v=0.1$, (b) $v=0.3$, (c) $v=0.5$.
The lines are fits to the algebraic function in Eq.~(\ref{eq:gapextrapolationscheme}).
The fits in Fig.~\ref{fig:Delta2finite}
are seen to agree very well with the data, showing a steep decrease
of the finite-size gap as a function of inverse system size. This indicates that
large system sizes are required to obtain reasonable gap extrapolations.

\begin{figure}[ht]
  \includegraphics[width=8cm]{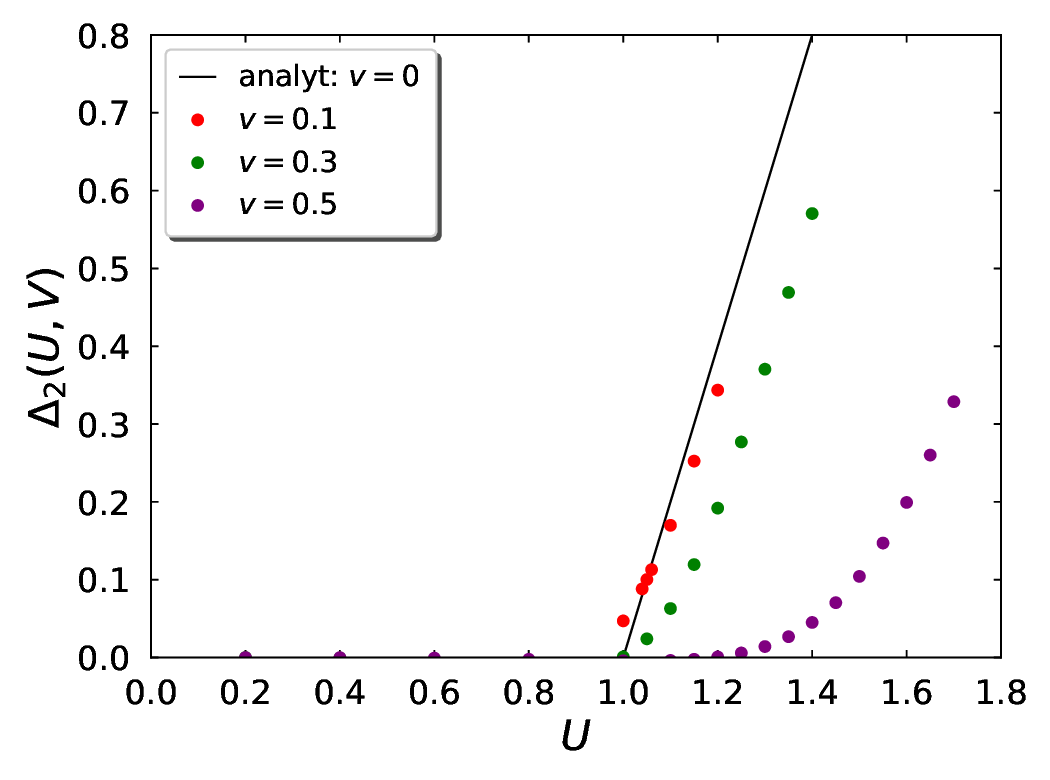}
  \caption{Two-particle gap $\Delta_2(U,V)$ for the extended $1/r$-Hubbard model
    as a function of $U$  for $v=0.1$ (red dots),
    $v=0.3$ (green dots), $v=0.5$ (purple dots),
    extrapolated from finite-size data with up to $L=64$ sites.
    The continuous line is the exact result in the thermodynamic limit
    for $V=0$, $\Delta_2(U,V= 0)= 2(U-1)$, see Eq.~(\ref{eq:Delta2TDLVzero}).\label{fig:Delta2}}
\end{figure}

The extrapolated gaps becomes {\em smaller\/}
as a function of~$V$, i.e.,
the nearest-neighbor interaction {\em reduces\/} the tendency to
form a Mott-Hubbard insulator.
The extrapolated gaps $\Delta(U,V)$ are shown in Fig.~\ref{fig:Delta2}
as a function of~$U$ for $v=0$, $v=0.1$, $v=0.3$, and $v=0.5$.
Apparently, the nearest-neighbor interaction not only shifts the
critical interaction to higher values, it also reduces
the size of the gap in the Mott insulating phase. 

At first sight, the {\em increase\/}
of the critical interaction is counter-intuitive
because one might argue that an additional repulsive nearest-neighbor
Coulomb interaction should favor the insulating state,
not the metallic state. From 
a wave-mechanical viewpoint, however, the repulsive nearest-neighbor interaction 
softens the two-particle scattering potential.
Figuratively speaking, particles
that are scattered by the weaker nearest-neighbor interaction~$V$
do not experience the stronger on-site interaction~$U$.
For a quantitative analysis, see Sect.~\ref{sec:MTnearest-neighbor}.

When $v= V/U$ is small, the change in the critical interaction strength is also small, and one
might think of using perturbation theory around the bare $1/r$-Hubbard model.
To test this idea, we consider
\begin{equation}
C(L;U,V)= \frac{e_0(L,L;U,V)-e_0(L,L;U,V= 0)}{V} \; .
\label{eq:DefCfrome0}
\end{equation}
In the limit $V\to 0$, leading-order perturbation theory gives
\begin{equation}
\lim_{V\to 0}C(L;U,V)= C^{\rm NN}(r= 1,L;U,V=0) \; ,
\label{eq:relationCandCNN}
\end{equation}
where $C^{\rm NN}(r= 1,L;U,V=0)$ is the nearest-neighbor density-density
correlation function at half band-filling for
the bare $1/r$-Hubbard model at finite system sizes~$L$,
see Eq.~(\ref{eq:CNNdef}).
As an example, for $v=0.3$ and $U\lesssim 0.7$, we find that
$C^{\rm NN}(r= 1,L;U,V=0)$ agrees fairly well with 
$C(L;U,V)$ from Eq.~(\ref{eq:DefCfrome0}). Around the Mott transition, however, the corrections become sizable,
more noticeably for larger systems.
Therefore, low-order perturbation theory around the limit $V= 0$
cannot be used to determine the 
critical interaction strength $U_{\rm c}(V)$ reliably.

\subsection{Momentum distribution}

\begin{figure}[t]
(a)  \includegraphics[width=7cm]{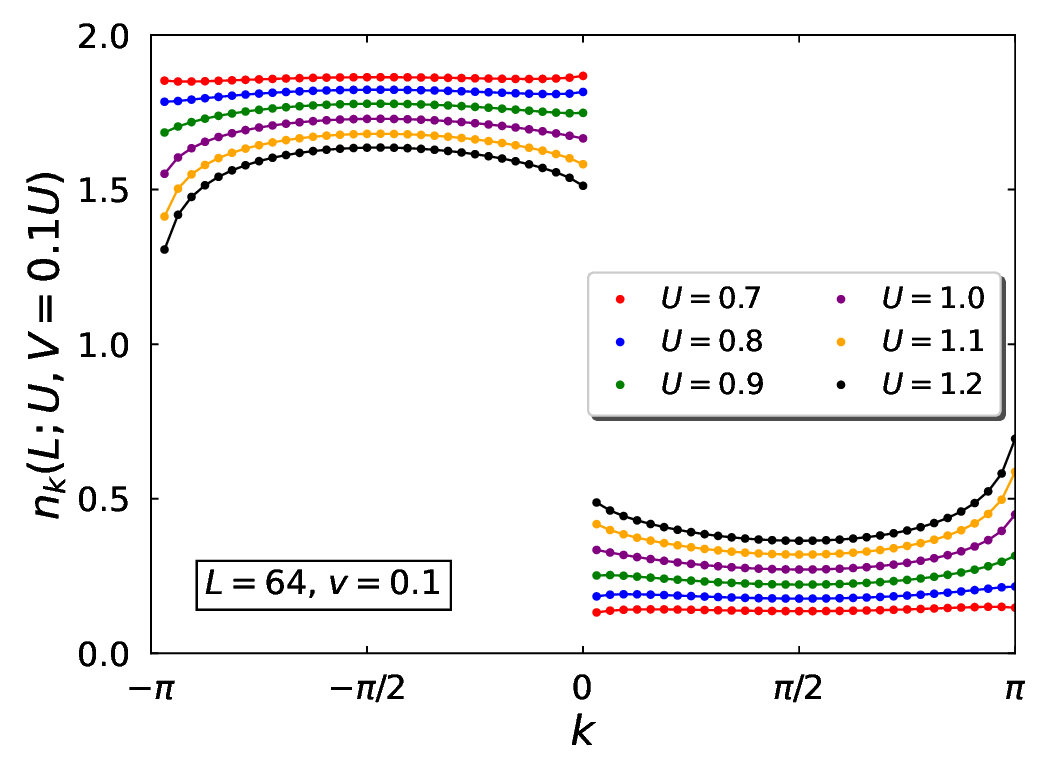}\\[12pt]
(b)  \includegraphics[width=7cm]{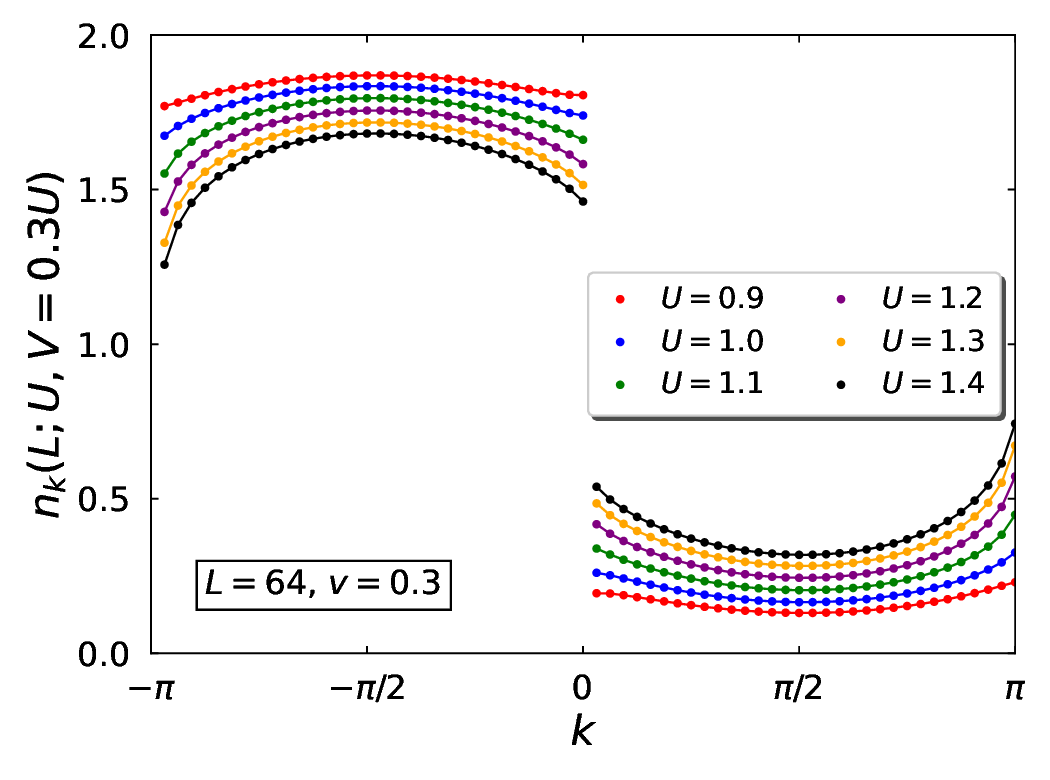}\\[12pt]
(c)  \includegraphics[width=7cm]{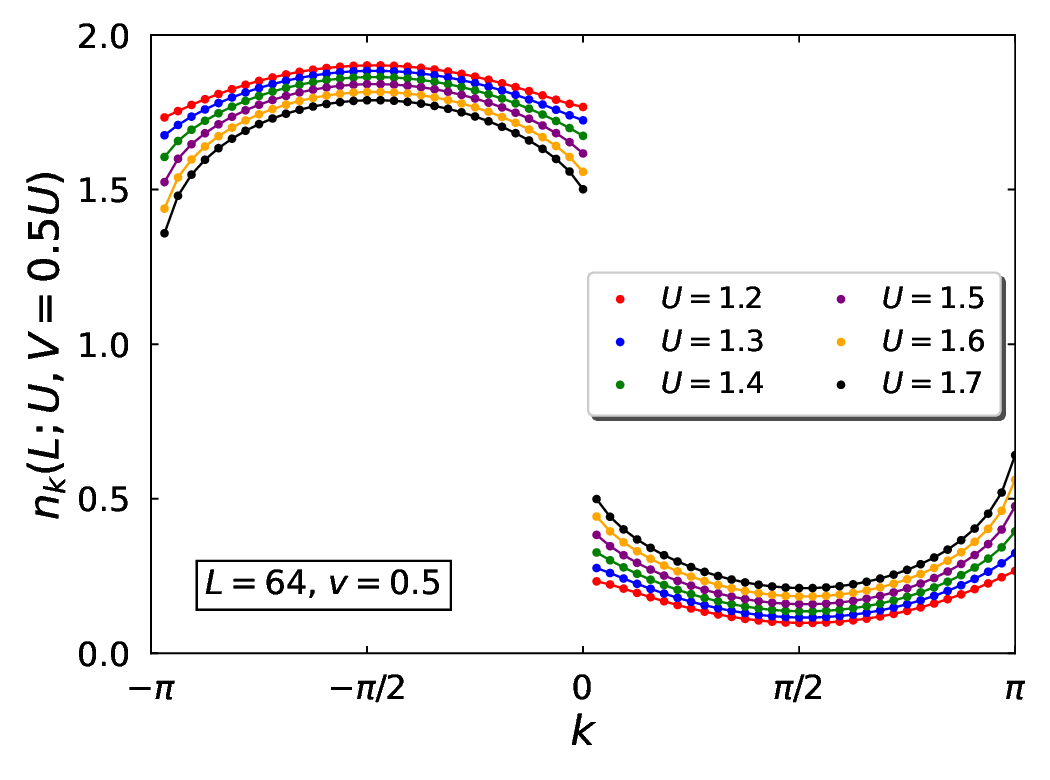}
 \caption{Momentum distribution $n_k(L;U,V)$ from DMRG at half band-filling
   for the extended $1/r$-Hubbard model
   for $L=64$ sites
   and  for various values for $U$
   for $v=0.1$, $v=0.3$, and $v=0.5$
   (from top to bottom).\label{fig:momdis}}
\end{figure}

In Fig.~\ref{fig:momdis} we show the momentum distribution from DMRG
at half band-filling for $L= 64$
sites for various values of~$U$ and $v= 0.1$, $v= 0.3$,
and $v= 0.5$ (from top to bottom). For small interactions, the
momentum distribution resembles that of a Fermi liquid with
all states $-\pi <k<0$ occupied and all states $0<k<\pi$ empty.
For small~$U$, 
low-energy scattering processes are limited to 
the vicinity of the sole Fermi point $k_{\rm F}= 0$.
Indeed, in the field-theoretical limit, $U,V \ll 1$, the model reduces
to a bare $g_4$-model of only right-moving particles~\cite{Thierrybook}.
This `non-interacting Luttinger liquid' displays a jump discontinuity at
$k_{\rm F}$.

However, the $1/r$-Hubbard model is defined on a lattice
and the bandwidth is finite. Consequently, the second Fermi point at
$k_{{\rm F},2}= -\pi$ starts to play a role when $U$ becomes
large, of the order of half the bandwidth.
States near $k_{{\rm F},2}$
are depleted more quickly as a function of~$U$ than those deeper in the
Brillouin zone. Therefore, as seen in Fig.~\ref{fig:momdis},
the momentum distribution develops a maximum around $k= -\pi/2$,
with a corresponding minimum around $k= \pi/2$.

These considerations show that the Luttinger parameter must deviate from
unity, $K(U,V)< 1$, for all $(U,V)$, even though corrections to unity
are (exponentially) small for $U,V \ll 1$.
Therefore, the momentum distribution is a continuous function
in the (extended) $1/r$-Hubbard model for all $U,V>0$.

In contrast to the case $V= 0$~\cite{GebhardLegezaOneoverrHM},
there is no Fano resonance discernible
in the slope of the momentum distribution at $k= -\pi$
as the slope is always positive at $k= -\pi$.
This indicates that the bound state for $V=0$ moves away from the band edge
for finite $V>0$ and thus cannot be detected in the momentum distribution.
Consequently, we cannot use the resonance to locate
the metal-insulator transition in the extended $1/r$-Hubbard model.

\subsection{Structure factor and CDW order parameter}

\begin{figure}[t]
(a)  \includegraphics[width=7cm]{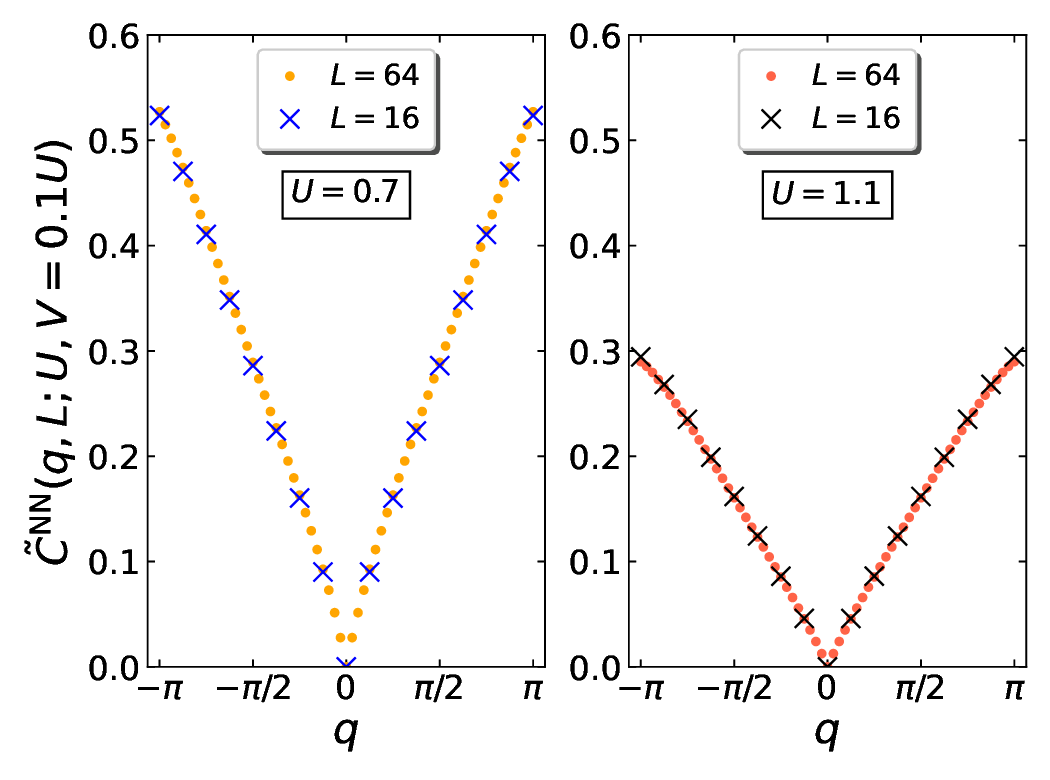}\\[12pt]
(b)  \includegraphics[width=7cm]{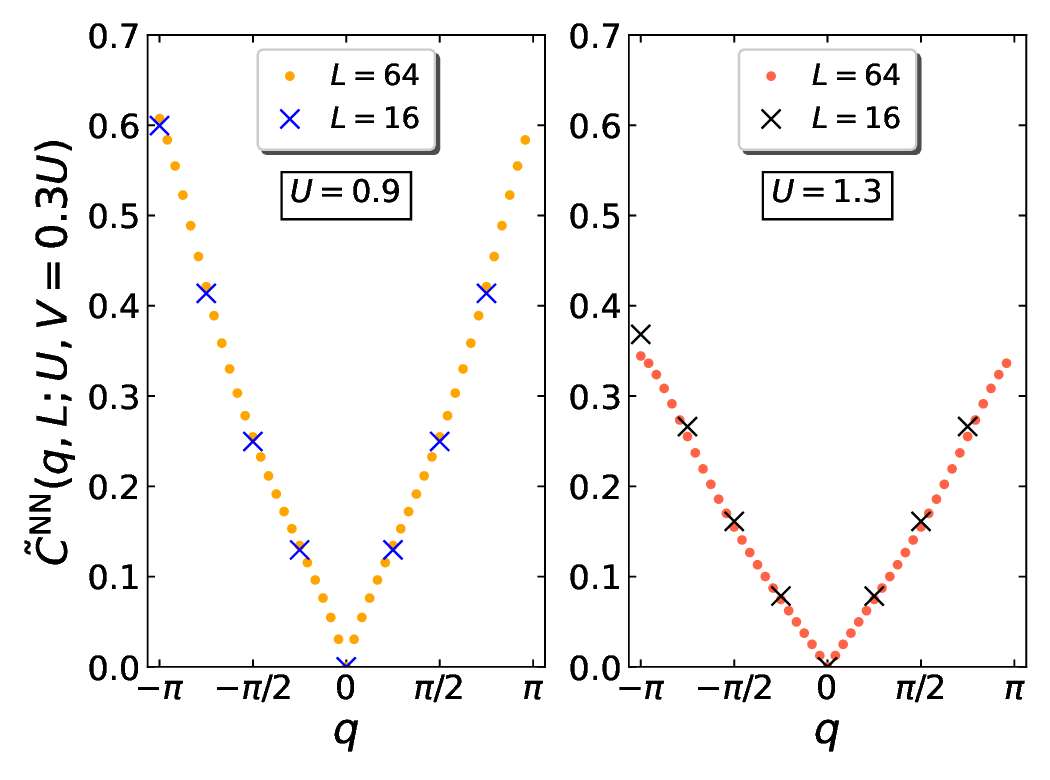}\\[12pt]
(c)  \includegraphics[width=7cm]{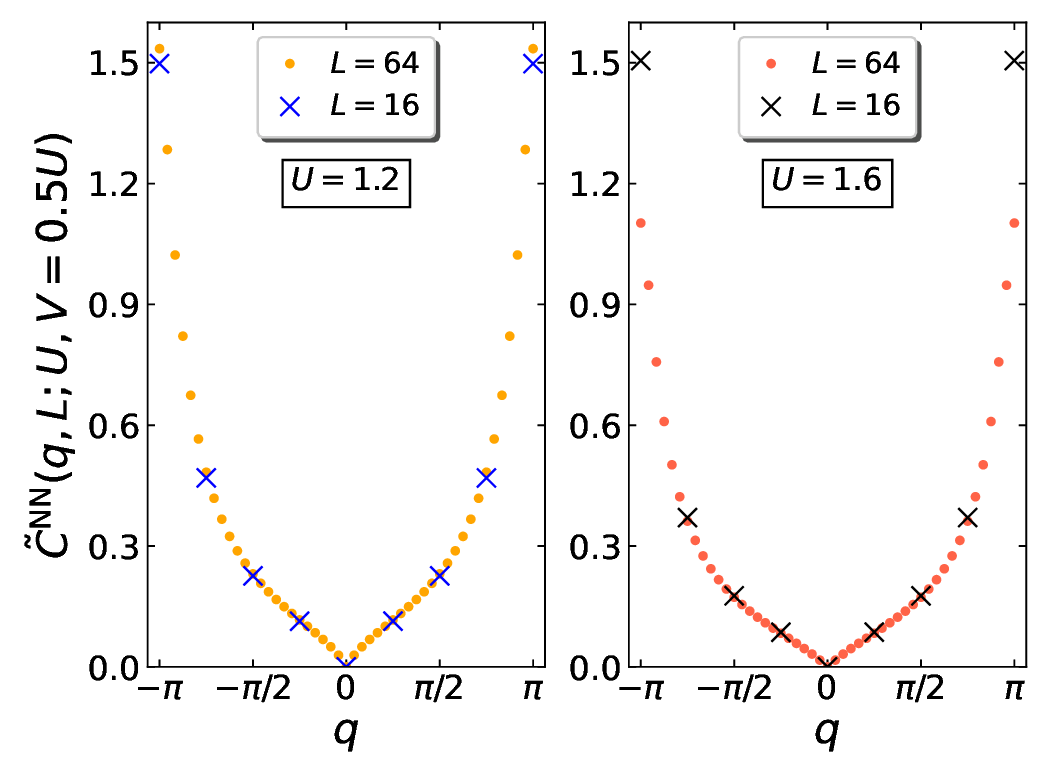}
 \caption{Structure factor $\tilde{C}^{\rm NN}(q,L;U,V)$
   for the extended $1/r$-Hubbard model
    for $L=16, 64$ below (left) and above (right) the Mott transition
    for (a) $v=0.1$, (b) $v=0.3$, (c) $v=0.5$.\label{fig:CNNq}}
\end{figure}

Lastly, we show the structure factor from DMRG in Fig.~\ref{fig:CNNq}
for $v=0.1$, $v=0.3$, and $v=0.5$ (from top to bottom)
for the extended $1/r$-Hubbard model at system sizes $L=16,64$
below (left) and above (right) the Mott transition.
It is seen that the finite-size effects are fairly small but larger
systems permit
a much better resolution in momentum space.
In comparison with the exact result for the non-interacting system,
\begin{equation}
\tilde{C}^{\rm NN}(q,n=1;U=0,V=0)=\frac{|q|}{\pi} \; ,
\end{equation}
we see that the local interaction reduces the charge fluctuations.
This is expected because the suppression of double occupancies likewise
reduces the number of holes and the charges are more homogeneously distributed 
in the system. Therefore, the charge correlations become smaller when
we compare the left and right figure in the same row.

The nearest-neighbor interaction counters the effect of the Hubbard interaction
because nearest-neighbor pairs of a double occupancy
and a hole are energetically
favorable. Therefore, the charge correlations increase when we
go from top to bottom in the left/right row, even though $U$ also increases
from top to bottom.

\begin{figure}[t]
  \begin{center}
   (a) \includegraphics[width=8cm]{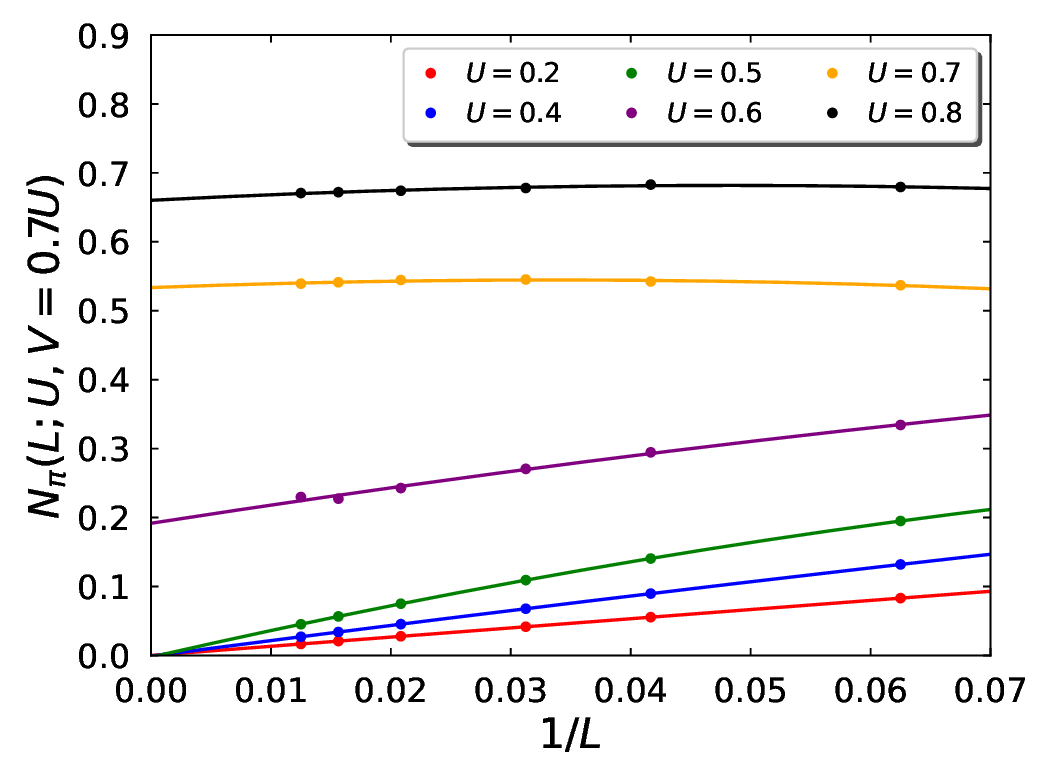}\\
   (b) \includegraphics[width=8cm]{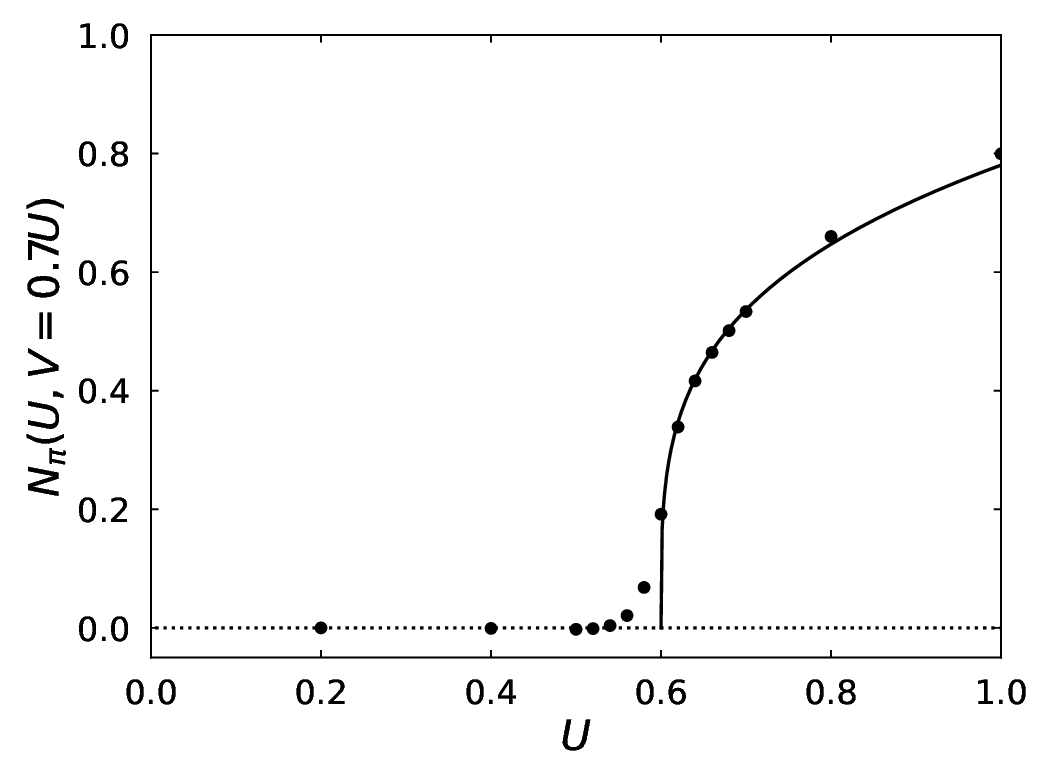}
  \end{center}
 \caption{(a) CDW order parameter $N_{\pi}(L;U,V)$ for the extended half-filled $1/r$-Hubbard model
    as a function of $1/L$ ($L\leq 80$) for $v=0.7$ and various $U$-values. Lines are a second-order polynomial fit in $1/L$,
    see Eq.~(\ref{eq:CDWfit2nd});
    (b) Extrapolated CDW order parameter $N_{\pi}(U,V=0.7U)$ as a function of $U$. The line is an algebraic fit to the data in the vicinity
    of the CDW transition, see Eq.~(\ref{eq:CDWexponent}), with $U_{\rm c}(v=0.7)=0.6$, $N_0=1$ and $2\nu=0.3$.\label{fig:CDWorderparameter}}
\end{figure}

When the nearest-neighbor interaction increases beyond a certain threshold value
$V_{\rm c}(U)$, the ground state displays charge-density-wave order.
In Fig.~\ref{fig:CDWorderparameter}(a) we show the charge-density-wave order parameter $N_{\pi}(L;U,V=0.7U)$, see Eq.~(\ref{eq:ourCDWorderparameterdef}), 
as a function of $1/L$ for various values of~$U$, and 
the extrapolated result $N_{\pi}(U,V=0.7U)$ into the thermodynamic limit using a second-order polynomial fit
in Fig.~\ref{fig:CDWorderparameter}(b),
\begin{equation}
    N_{\pi}(L;U,V) = N_{\pi}(U,V) +\frac{N_1(U,V)}{L} + \frac{N_2(U,V)}{L^2} \; .
    \label{eq:CDWfit2nd}
\end{equation}
Apparently, the CDW order parameter is continuous over the CDW transition.
Close to the transition, $U\gtrsim U_{\rm c}(V)$,
\begin{equation}
N_{\pi}(U,V) =N_0\left[U-U_{\rm c}(V)\right]^{2\nu} \; ,
\label{eq:CDWexponent}
\end{equation}
where $\nu$ is the critical exponent for the CDW order parameter $D(U,V)$.
Note that we pass the CDW transition for a fixed ratio $v=U/V$. 

To make use of Eq.~(\ref{eq:CDWexponent}), 
the critical interaction $U_{\rm c}(V)$ must be known.
In addition, the region of validity of Eq.~(\ref{eq:CDWexponent}) is unknown {\sl a priori}.
Typically, one has to study system parameters close to the transition to obtain
a reliable estimate for $\nu$. Therefore, very large system sizes might be necessary 
to reach the scaling limit, and we have to be satisfied with the result from Fig.~\ref{fig:CDWorderparameter}(b)
that the CDW transition at $v=0.7$ is continuous with exponent $\nu \leq 1/2$. 

\section{Mott transition}
\label{sec:MTnearest-neighbor}

In this section we determine the critical value for the Mott transition in the extended $1/r$-Hubbard model.
We investigate the two-particle gap, the ground-state energy, the Luttinger parameter, and
the structure factor at the Brillouin zone boundary to locate the critical interaction strength $U_{\rm c}(V)$.
The Mott transition remains continuous for
all $V/U$.

\subsection{Two-particle gap}
\label{subsec:tpgapgamma2}

In our previous work~\cite{GebhardLegezaOneoverrHM}, we showed that the
exponent $\gamma_2(U)=\gamma_2(U,V=0)$ sensitively depends on~$U$ in the vicinity
of the Mott-Hubbard transition, and  the
critical interaction for the $1/r$-Hubbard model, $U_{\rm c}(V=0)=1$,
was obtained with an accuracy of one per mil.

\begin{figure}[t]
  \begin{center}
    \includegraphics[width=8cm]{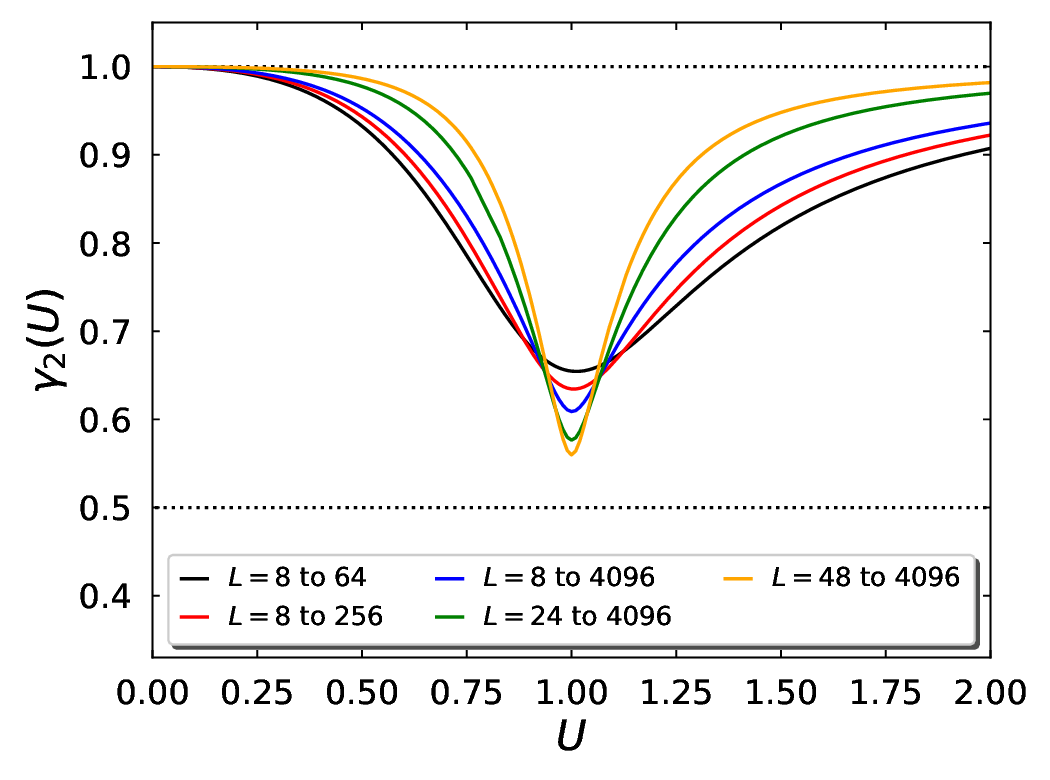}
  \end{center}
  \caption{Extrapolated gap exponent $\gamma_2(U)=\gamma_2(U,V=0)$ using the analytical expression of the two-particle gap in Eq.~(\ref{eq:Delta2analyt}). Various system sizes are used, ranging from $L=8,16,24,32,48,64,80,96,128,256,512,1024,2048,4096$.\label{fig:analytgammaofUforV0}}
\end{figure}

To illustrate this result for the bare $1/r$-Hubbard model,   
in Fig.~\ref{fig:analytgammaofUforV0}
we show the extrapolated gap exponent $\gamma_2(U)\equiv \gamma_2(U,V=0)$ using the analytic expression~(\ref{eq:Delta2analyt}) 
for various combinations of system sizes in the range $L=8,16,24,32,48,64,80,96,128,256,512,1024,2048,4096$. 

The minimal value for $\gamma_2(U)$ depends on the selected range of system sizes. 
The gap exponent in the thermodynamic limit, see Eq.~(\ref{eq:gamma2TDL}), cannot be reproduced from finite-size studies but it is approached systematically with increasing system size. 
Furthermore, it can be seen from Fig.~\ref{fig:analytgammaofUforV0} that the inclusion of smaller system sizes such as $L=8,16$ leads to stronger deviations so that the smallest system sizes should be discarded. 
Note, however, that the {\em position\/} of the minimum and thus the critical interaction strength are very well reproduced in all cases.
Therefore, the minimum of $\gamma_2(U,V)$ permits to locate the Mott transition $U_{\rm c}(V)$ fairly accurately.

\begin{figure}[b]
  \begin{center}
    \includegraphics[width=8cm]{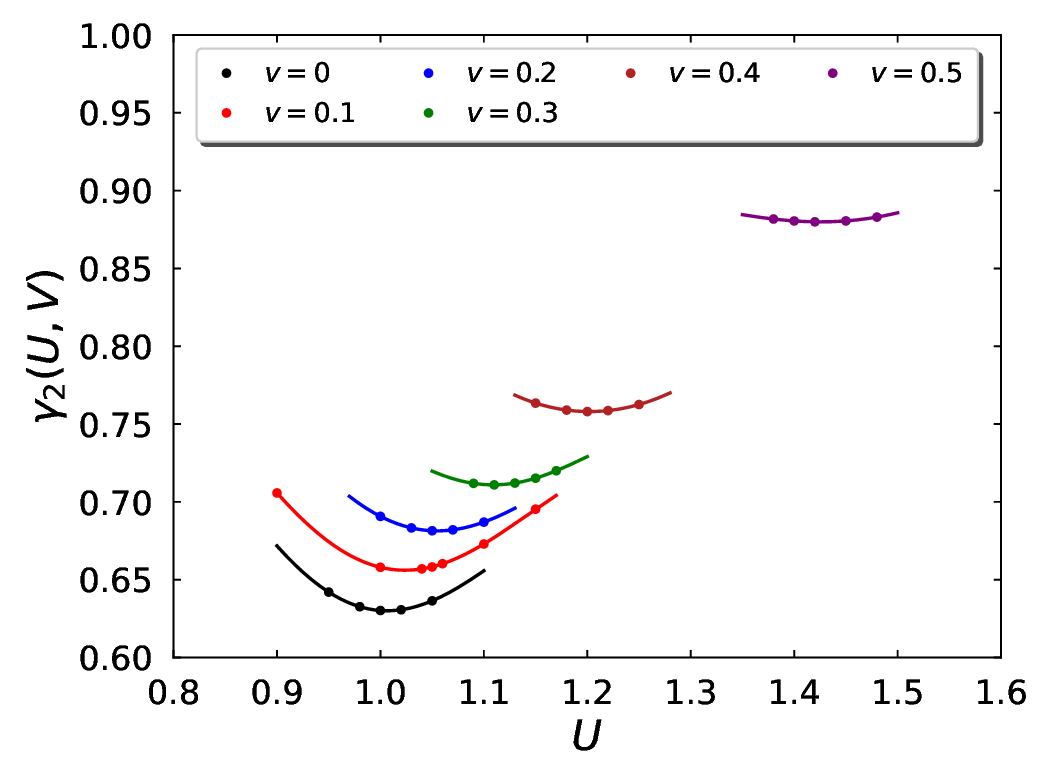}
  \end{center}
  \caption{Exponent $\gamma_2(U,V)$ 
    for the two-particle gap in the extended $1/r$-Hub\-bard model as a function of $U$ for various values of $v=V/U$,
    based on system sizes $16\leq L \leq 80$.
The minimum of the curve determines $U_{\rm c,gap}(V)$.\label{fig:gammaofUinHM}}
\end{figure}

In Fig.~\ref{fig:gammaofUinHM} we display the exponent
$\gamma_2(U,V)$, as obtained from the fit of the finite-size data
in the range $16\leq L \leq 80$ to the algebraic function in Eq.~(\ref{eq:gapextrapolationscheme}).
Also shown in the figure are the quartic fits around the minima which lead to
the critical interactions $U_{\rm c,gap}(V)$ listed in table~\ref{tab:one}.
Note that the curves flatten out for increasing $v$ so that it becomes more difficult to determine accurately the minima for $v\to 0.5$.

\begin{table}[t]
  \begin{ruledtabular}
    \begin{tabular}[t]{rrrrrr}
      $V/U$ & $U_{\rm c,gap} (V)$ & $U_{\rm c,gs}(V)$ & $U_{\rm c,LL}(V)$ & $U_{\rm c,sf}(V)$ & $\overline{U}_{\rm c}(V)$ \\
      \hline \\[-7pt]
      0   & 1.009 & 1.000 & 1.033 & 0.965 &  1.002 \\
      0.1 & 1.024 & 1.022 & 1.056 & 0.984 &  1.021 \\
      0.2 & 1.055 & 1.056 & 1.090 & 1.018 &  1.055 \\
      0.3 & 1.109 & 1.116 & 1.144 & 1.075 &  1.111 \\
      0.4 & 1.202 & 1.221 & 1.243 & 1.175 &  1.210 \\
      0.5 & 1.425 & 1.500 & 1.540 & 1.456 &  1.480\\
      0.6 & 0.828 & 0.838 & 0.883 & 0.876 &  0.856 \\
      0.7 & 0.587 & 0.600 & 0.616 & 0.611 &  0.604 \\
    \end{tabular}
          \end{ruledtabular}
  \caption{Critical interaction strengths for the extended $1/r$-Hubbard model,
as obtained from the two-particle gap, the ground-state energy, the Luttinger parameter,
and the structure factor for systems with $16\leq L\leq 80$ lattice sites.
    For $V=0$, the exact
    result in the
    thermodynamic is known~\cite{GebhardRuckenstein},
    $U_{\rm c}(V= 0)= 1$.\label{tab:one}}
  \end{table}

The comparison with the exact value for $V=0$ shows that the gap exponent $\gamma_2(U,V)$ provides a fairly accurate estimate for
the critical interaction. The same accuracy can be obtained when using the ground-state energy exponent $\gamma_0(U,V)$,
as we shall show next.

\begin{figure}[b]
  \begin{center}
    \includegraphics[width=8cm]{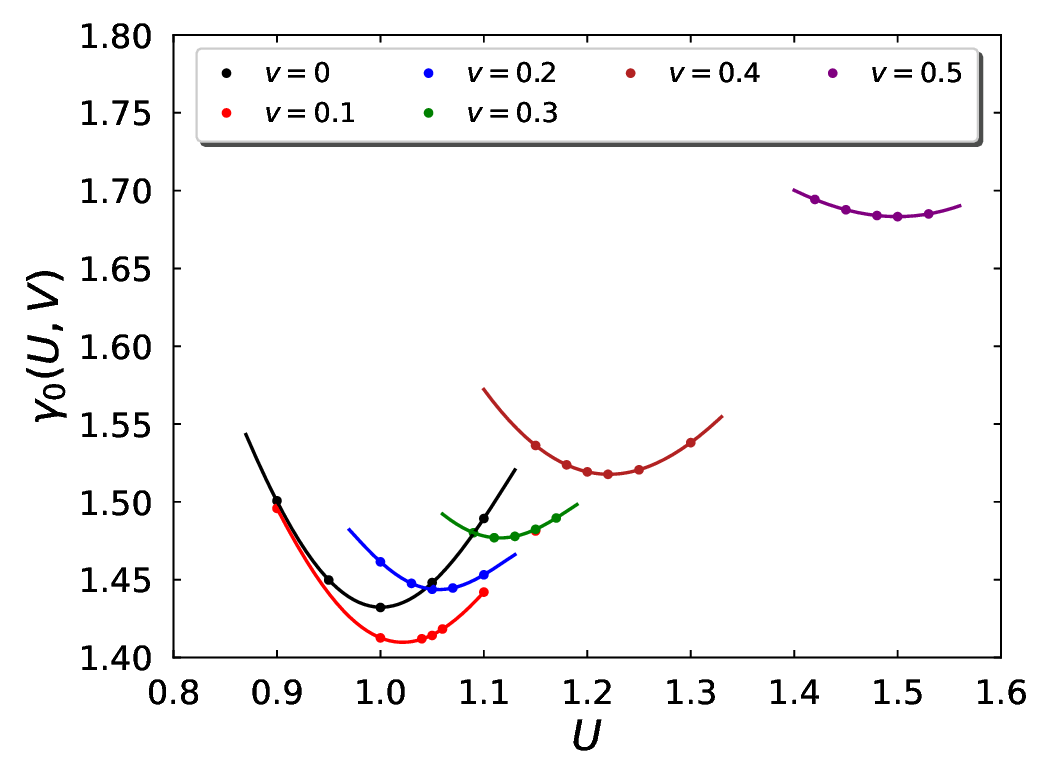}
  \end{center}
  \caption{Exponent $\gamma_0(U,V)$ 
    for the ground-state energy of the extended $1/r$-Hub\-bard model as a function of $U$ for various values of $v=V/U$,
    based on system sizes $16\leq L \leq 80$.
The minimum of the curve determines $U_{\rm c,gs}(V)$.\label{fig:groundstategammaofUinHM}}
\end{figure}

\subsection{Ground-state energy}
\label{subsec:gsenergygamma0}

As seen from Eq.~(\ref{eq:gamma0TDL}), the $1/L$ corrections to the ground-state energy density also permit to locate the Mott transition in the extended $1/r$-Hubbard model,
in the same way as the two-particle gap. 
In Fig.~\ref{fig:groundstategammaofUinHM} we show the exponent
$\gamma_0(U,V)$, as obtained from the fit of the finite-size data
in the range $16\leq L \leq 80$ to the algebraic function in Eq.~(\ref{eq:ezeroextrapolation}).
Also shown in the figure are the quartic fits around the minima which lead to
the critical interactions $U_{\rm c,gs}(V)$ listed in table~\ref{tab:one}.

The critical interaction strengths obtained from the minima of $\gamma_0(U,V)$ very well agree with the exact result at $V=0$ and with the values obtained 
from the gap exponent $\gamma_2(U,V)$ with deviations in the low percentage range. 
Therefore, we can be confident that 
we found reliable estimates for the critical interaction strength for the Mott transition.

\subsection{Luttinger parameter}
\label{subsec:Luttingerparameter}

As an alternative to locate the Mott transition, we monitor the Luttinger
parameter and determine $U_{\rm c}(U,V)$ from the condition~\cite{Thierrybook}
\begin{equation}
  K(U_{\rm c}(V),v)= \frac{1}{2}
  \label{eq:Kmustbeonehalf}
\end{equation}
for fixed ratios $v= V/U$, see also Ref.~\cite{GebhardLegezaOneoverrHM}.

\begin{figure}[b]
  \begin{center}
    \includegraphics[width=8cm]{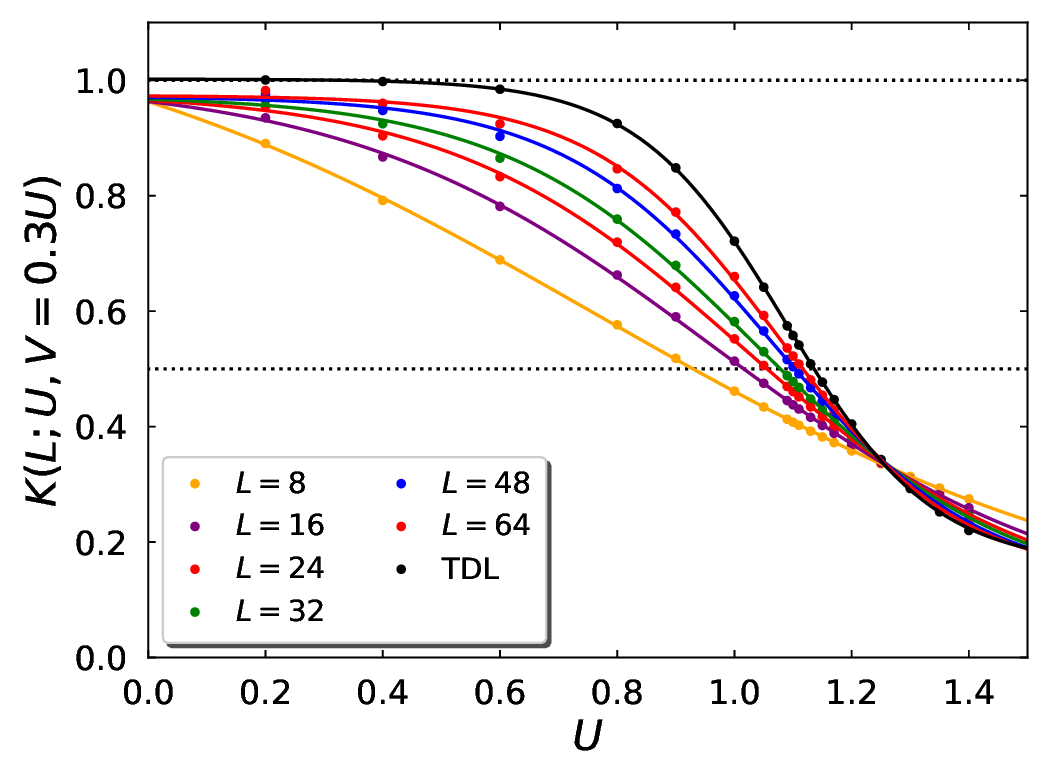}
  \end{center}
  \caption{Luttinger parameter $K(L;U,V)$ from DMRG
 for the extended $1/r$-Hubbard model with nearest-neighbor interaction $V= 0.3U$
    as a function of $U$ for system sizes $L= 8,16,24,32,48,64$
    including a second-order polynomial extrapolation
    to the thermodynamic limit.
    The intersection of the extrapolation with $K_{\rm c}= 1/2$ determines
    $U_{\rm c}(V)$.\label{fig:LuttParameter03}}
\end{figure}

In Fig.~\ref{fig:LuttParameter03}
we show the Luttinger parameter
$K(L;U,V)$ from DMRG
 for the extended $1/r$-Hubbard model with nearest-neighbor interaction $V= 0.3U$
    as a function of $U$ for system sizes $L= 8,16,24,32,48,64$
    including a second-order polynomial extrapolation
    to the thermodynamic limit.
    The intersection of the extrapolation into the thermodynamic limit
    with $K_{\rm c}= 1/2$ determines
    $U_{\rm c}(V)$. To obtain a reliable estimate
    for the intersection we can either
    use the two data points closest to the transition
    and perform a linear interpolation, in this case $U= 1.1$
    and $U= 1.2$. Alternatively, we use a four-parameter fit of the whole
    data set that employs the information that the Luttinger parameter
    deviates from unity by exponentially small terms for $U,V\to 0$.
    Thus, we use
    \begin{equation}
      K(U,V)= a + b\tanh(c+ dU)
      \label{eq:fitMTDLforallU}
    \end{equation}
    to fit the extrapolated data for finite values of $U$ to a continuous
    curve which is parameterized by $a,b,c,d$ that depend on~$v$.
    Then, we solve Eq.~(\ref{eq:Kmustbeonehalf}) for $U_{\rm c,LL}(V)$.
    The results are also listed in table~\ref{tab:one}.
  
    Alternatively, we could have solved Eq.~(\ref{eq:Kmustbeonehalf}) for each system size, and extrapolated the resulting
    system-size dependent critical interactions strengths to the thermodynamic limit.
    Since the results deviate more strongly from the exact value for $V=0$, we refrain from pursuing this approach further.
    
As seen from table~\ref{tab:one}, the critical value from Luttinger parameter systematically
overestimate the correct interaction strengths by some three percent. A similar effect was found for
the charge-density-wave transition in a one-dimensional model for
spinless fermions with nearest-neighbor interactions (`$t$-$V$-model')~\cite{PhysRevB.106.205133}.
Apparently, much larger systems are required to overcome this systematic error.
In this work, we do not apply correction factors for a better fit but use the critical interaction strengths
$U_{\rm c,LL}(V)$ as an upper bound to the exact value $U_{\rm c}(V)$.

\begin{figure}[b]
  \begin{center}
    \includegraphics[width=8cm]{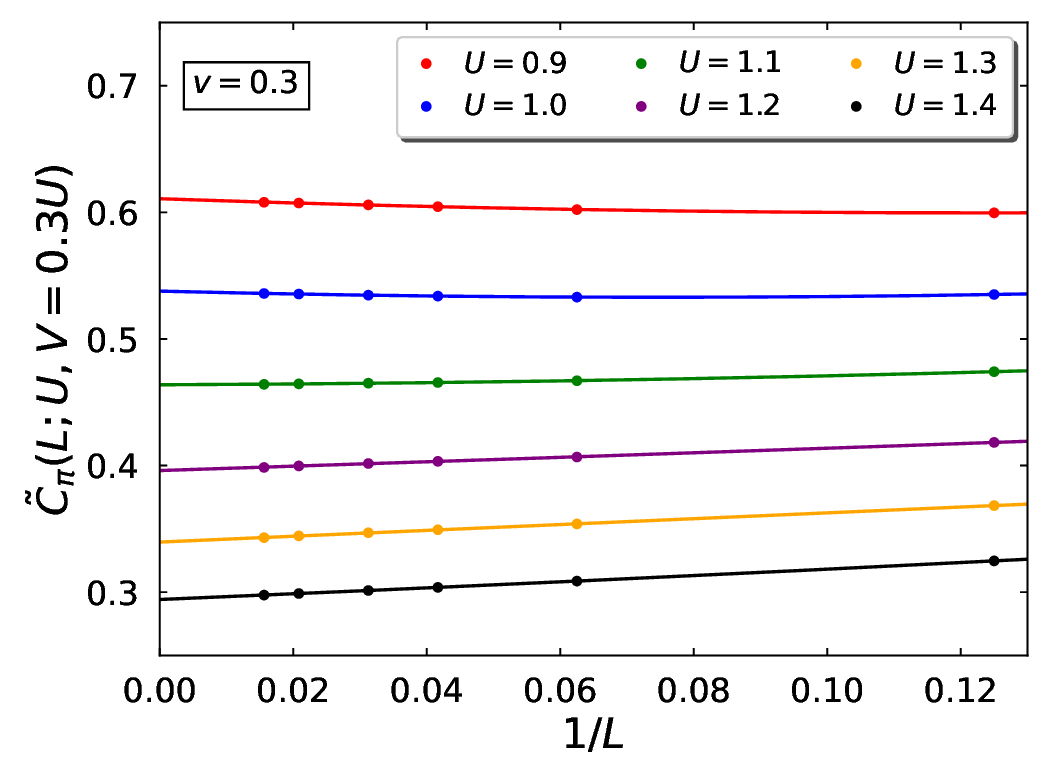}
  \end{center}
  \caption{Structure factor $\tilde{C}_{\pi}(L;U,V)$ at $q=\pi$ as a function of $1/L$ for various values of $U$ 
  for the extended $1/r$-Hubbard model with nearest-neighbor interaction $V= 0.3U$
for system sizes $L= 8,16,24,32,48,64$. Lines are a second-order polynomial extrapolation
    to the thermodynamic limit, see Eq.~(\ref{eq:fitstucturefactorinL}).\label{fig:structurefactoratpi03}}
\end{figure}

\subsection{Structure factor and CDW order parameter} 
\label{subsec:strucfactor}

For the $1/r$-Hubbard model, the finite-size corrections to the structure factor $\tilde{C}_{\pi}(U,V)\equiv \tilde{C}(\pi;U,V)$,
\begin{equation}
    \tilde{C}_{\pi}(L;U,V)= \tilde{C}_{\pi}(U,V) +\frac{C_1(U,V)}{L} + \frac{C_2(U,V)}{L^2} \; ,
    \label{eq:fitstucturefactorinL}
\end{equation}
and the CDW order parameter $N_{\pi}(L;U,V)$, see Eq.~(\ref{eq:CDWfit2nd}),
permit to locate the critical interaction strength. 
In Fig.~\ref{fig:structurefactoratpi03} we show the structure factor for $v=0.3$ and various values of $U$ as a function of inverse system size
for $L=8,16,24,32,48,64$. 

\begin{figure}[b]
  \begin{center}
  (a)  \includegraphics[width=8cm]{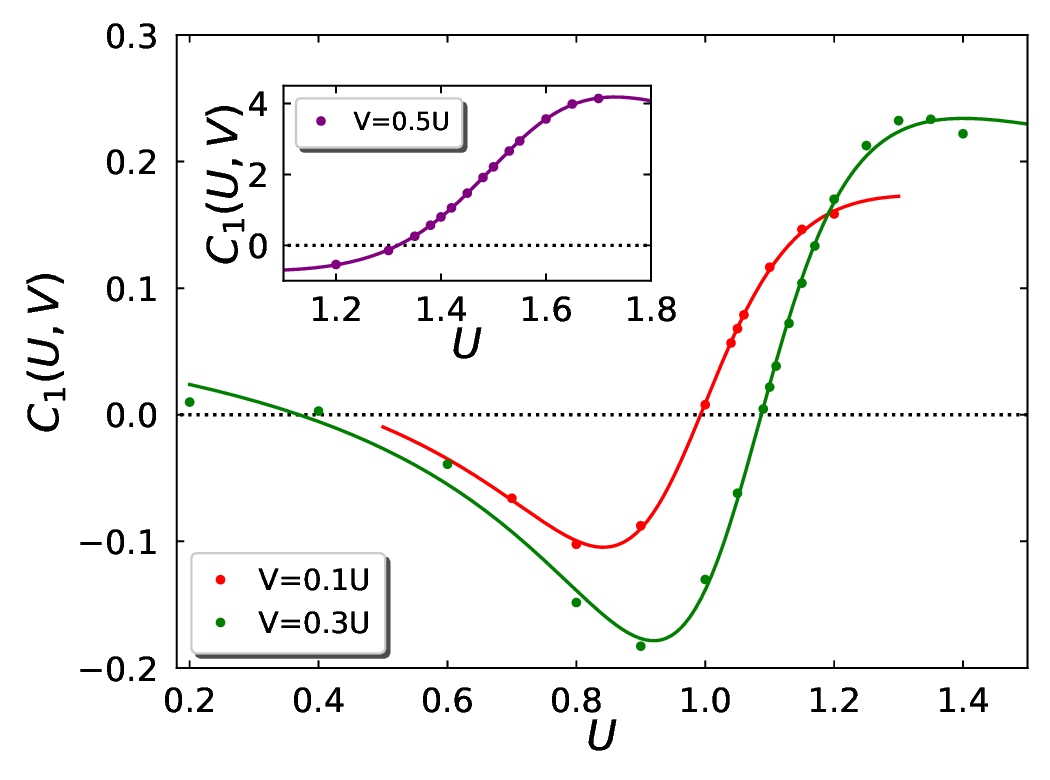}\\
  (b) \includegraphics[width=8cm]{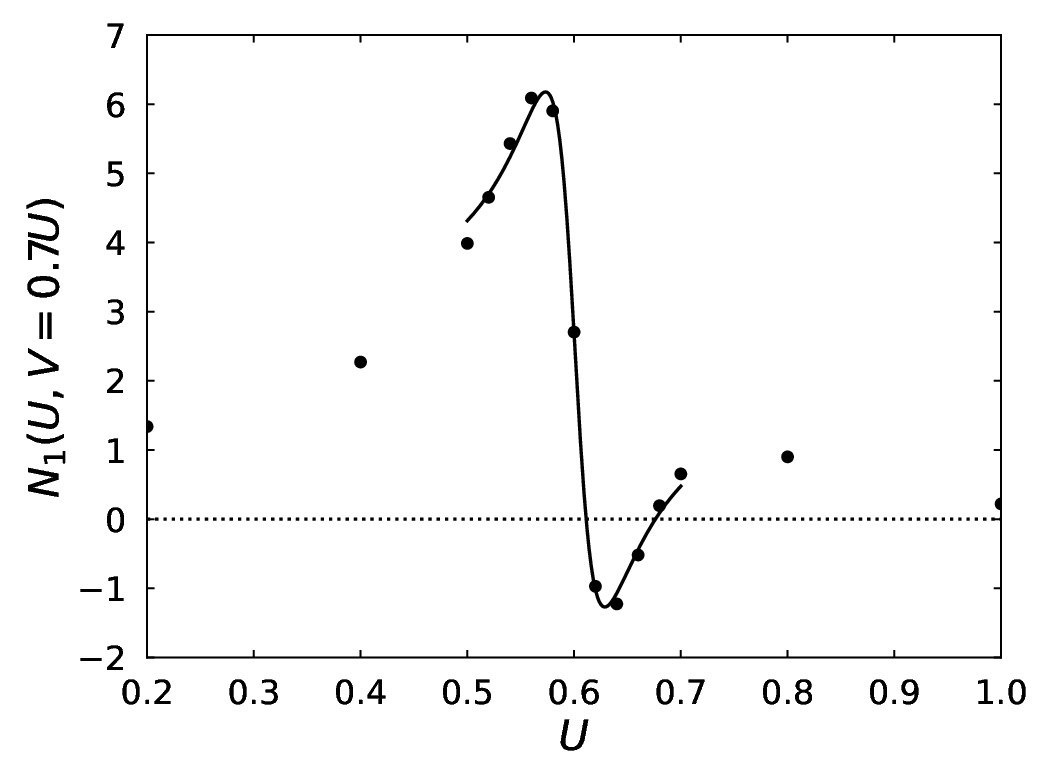}
  \end{center}
  \caption{(a) Finite-size coefficient $C_1(U,V)$ of the structure factor as a function of $U$
  for the extended $1/r$-Hubbard model for $v=0.1$ and $v=0.3$ (inset: $v=0.5$). 
  (b)  Finite-size coefficient $N_1(U,V=0.7U)$ for the CDW order parameter, see Eqs.~(\ref{eq:ourCDWorderparameterdef}) and~(\ref{eq:CDWfit2nd}). 
  Lines are fitted Fano resonance curves, see Eq.~(\ref{eq:Fanoformula}).\label{fig:structurefactorC1}}
\end{figure}

As can be seen from the figure, 
the coefficient in $1/L$ changes its sign at the critical
interaction strength, 
\begin{equation}
C_1(U_{\rm c,sf}(V),V)=0 \; .
\label{eq:C1iszero}
\end{equation}
To see this more clearly, in Fig.~\ref{fig:structurefactorC1}(a) we show the coefficient $C_1(U,V)$
as a function of $U$ for $v=0.1$, $v=0.3$, and $v=0.5$, and fit the data to 
a Fano resonance,
\begin{equation}
     C_1^{\rm Fano}(U,V)=a(V)  + b(V)
  \frac{[q_{\rm F}(V)\Gamma(V) +U-U_{\rm c}(V)]^2}{[\Gamma(V)]^2+[U-U_{\rm c}(V)]^2}\; .
    \label{eq:Fanoformula}
\end{equation}
Analogously, we find the critical interaction strengths in the CDW phase from 
the $1/L$ corrections to the CDW order parameter~(\ref{eq:ourCDWorderparameterdef}), see Eq.~(\ref{eq:CDWfit2nd}),
in Fig.~\ref{fig:structurefactorC1}(b).

As in our study of the $1/r$-Hubbard model~\cite{GebhardLegezaOneoverrHM},
a bound state that interacts with the continuum shows up in physical quantities and thus
contributes a Fano resonance to various physical quantities, with weight of the order $1/L$.
Using the Fano resonance formula and the conditions $C_1(U_{\rm c,sf},V)=0=N_1(U_{\rm c,sf},V)$, the $1/L$-corrections
of the structure factor and the CDW order parameter
provide the estimate $U_{\rm c,sf}(V)$ for the critical interaction.
The resulting data are listed for various~$v$ in table~\ref{tab:one}.

The critical interaction strength $U_{\rm c,sf}(V)$ systematically underestimates the exact value for the
Mott transition by a few percent. Together with the critical interaction strength from the Luttinger parameter $U_{\rm c,LL}(V)$
we thus can set tight limits to $U_{\rm c}(V)$.

\subsection{Critical interactions for fixed interaction ratios}
\label{subsec:critint}

In table~\ref{tab:one} we collect the results for the critical interaction strengths $U_{\rm c}(V)$
obtained from the analysis of the two-particle gap, the ground-state energy, the Luttinger parameter, and
the structure factor for $v=V/U=0,0.1,0.2,0.3,0.4,0.5,0.6,0.7$, as obtained from 
the previous four subsections. 
We observe that 
\begin{itemize}
\item[--] the arithmetic average of the four values, $\overline{U}_{\rm c}$, reproduces the exact result at $V=0$ 
with an accuracy of a few per mil;
\item[--] the values for $U_{\rm c,gs}(V)$ are close to the average for all $V$, with a deviation
below two percent. Therefore, the ground-state exponent alone provides a reliable estimate
for $U_{\rm c}(V)$ in all cases;
\item[--] the estimates $U_{\rm c,LL}(V)$, using the Luttinger parameter, and $U_{\rm c,sf}$,
using the structure factor, respectively, systematically overestimate and underestimate
the critical interaction strength for the transition from the Luttinger liquid to the Mott-Hubbard insulator. Therefore, they provide natural bounds to $U_{\rm c}(V)$ for $v\leq 0.5$;
\item[--] the transitions to the CDW insulator at $v=0.6, 0.7$ can be determined fairly accurately from 
all four approaches individually.
\end{itemize}
In Fig.~\ref{fig:phasediagram}, we connect the data points for $\overline{U}_{\rm c}(V)$
using a third-order spline interpolation. Error bars at the data points result from the 
overestimates and underestimates listed in table~\ref{tab:one}. 
In Fig.~\ref{fig:phasediagram} we also include the results from the analysis 
for the Mott transition between the Luttinger liquid and the CDW insulator at fixed $U=0.2$, 
as we discuss next.

\subsection{Transitions at fixed Hubbard interaction}
\label{subsec:fixedU}

Lastly, we study the metal-to-insulator transition at fixed Hubbard interaction~$U$
as a function of~$V$, namely for $U=0.2$ and $U=1.7$.

\subsubsection{Transition from Luttinger liquid to CDW insulator}
\label{subsubsec:smallU}

At $U=0.2$, we find a transition from the
Luttinger liquid metal to the CDW insulator at $V_{\rm c}(U=0.2)= 0.29\pm 0.01$.
The analysis follows the route outlined in the previous subsections, and will not be
repeated here. We increase $V$ in steps of $\Delta V=0.02$ around the transition.

Using the coefficient $\gamma_0$ from the ground-state energy, see Sect.~\ref{subsec:gsenergygamma0},
we find $V_{\rm c,gs}(U=0.2)=0.286$, the coefficient $\gamma_2$ from the two-particle gap
in Sect.~\ref{subsec:tpgapgamma2}
leads to $V_{\rm c,gap}(U=0.2)= 0.280$, and the Luttinger parameter 
of Sect.~\ref{subsec:Luttingerparameter} leads to
$V_{\rm c,LL}(U=0.2)=0.298$, almost identical to the values from the structure factor,
see Sect.~\ref{subsec:strucfactor}.
This leads to the average value quoted above.

Due to the absence of perfect nesting in the dispersion relation, it requires a finite interaction strength~$V$
to stabilize the CDW phase even at $U=0$. Qualitatively, Hartree-Fock theory leads to the same result.
Hartree-Fock theory systematically overestimates the stability
of the CDW phase and thus underestimates $V_{\rm c}(U)$, see Fig.~\ref{fig:phasediagram}.
The analytical approach can be improved by including second-order corrections to
Hartree-Fock theory, see, e.g., Refs.~\cite{PhysRevB.50.14016,PhysRevB.106.205133}.
This is beyond the purpose of our present analysis.

\subsubsection{Transition from Mott-Hubbard to CDW insulator}
\label{subsubsec:strongU}

\begin{figure}[t]
    \begin{center}
    (a) \includegraphics[width=6cm]{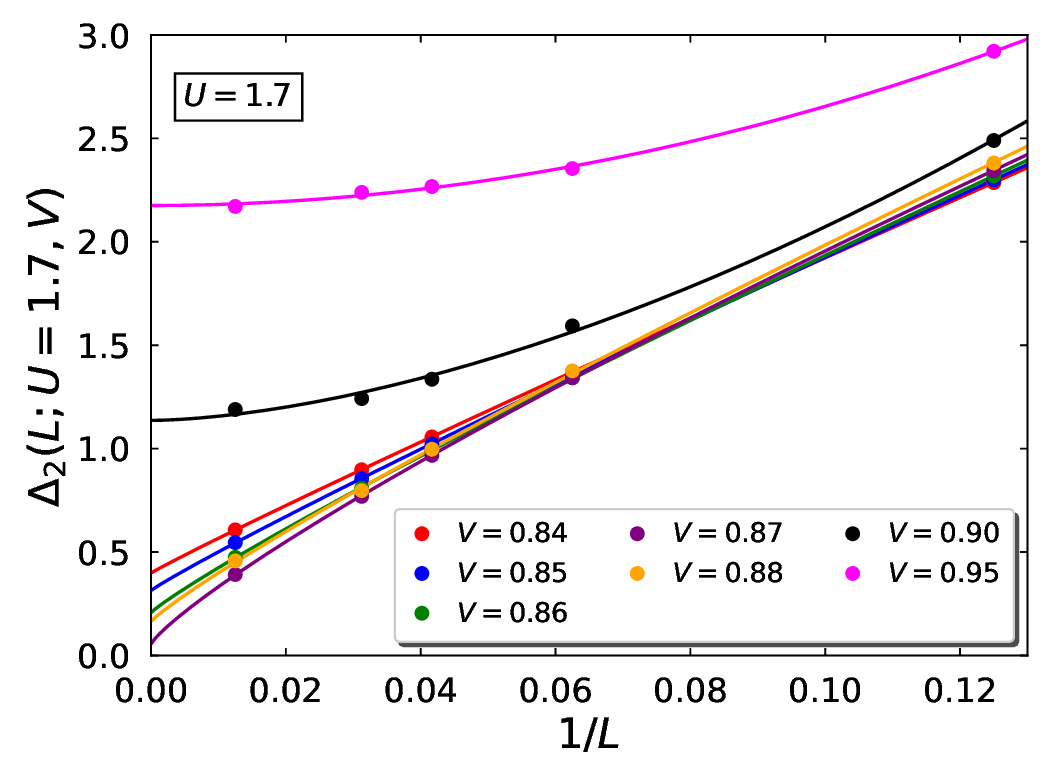}\\[6pt]
        (b) \includegraphics[width=6cm]{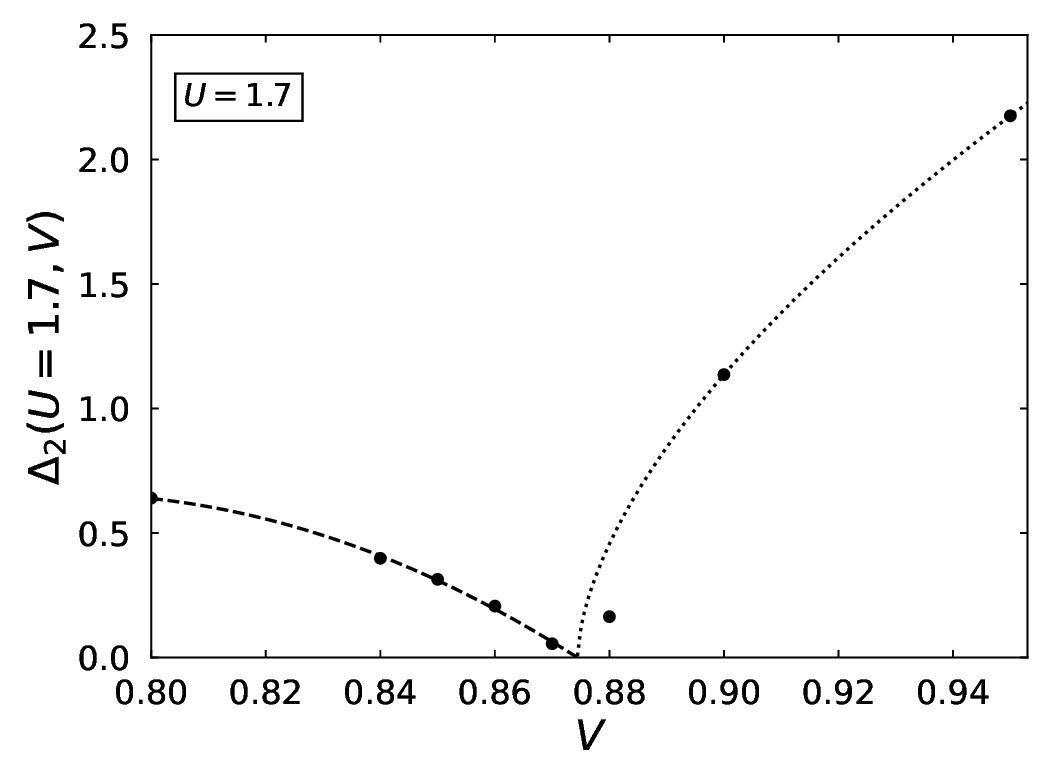}\\
    \end{center}
    \caption{(a) Two-particle gap for the extended $1/r$-Hubbard model at $U=1.7$ for various values of~$V$ as a function of $1/L$ for $L=8,16,32,64,80$.
   (b) Extrapolated two-particle gap as a function of~$V$.\label{fig:verticalU17}}
\end{figure}

At $U=1.7$, not included in the phase diagram in Fig.~\ref{fig:phasediagram},
we have a brief look at the transition from the Mott-Hubbard
insulator to the CDW insulator. 
The results for the two-particle gap are shown in Fig.~\ref{fig:verticalU17}.
They are corroborated by the behavior of the order parameter quantities
$C_{\pi}$ and $N_{\pi}$. The analysis of the parameters $\gamma_0$ and $\gamma_2$ lead to
quantitatively identical but less accurate results.

For the one-dimensional extended Hubbard model 
it is known that the critical interaction is larger than $U/2$. For the extended $1/r$-Hubbard model
we also find
$V_{\rm c}= 0.87\pm 0.01>1.7/2=0.85$ for the onset of the CDW. 
When expressed in units of the bandwidth,
the offset of $\delta_{\rm c}(U)=V_{\rm c}(U)-U/2$ agrees almost quantitatively with 
the value obtained from DMRG and QMC calculations for the one-dimensional
extended Hubbard model, $\delta_{\rm c}(U=1.7)\approx 0.02$, see Ref.~\cite{Jeckelmannreply2005}.

The shift $\delta_{\rm c}(U)$ can be determined analytically using higher-order strong-coupling
perturbation~\cite{PVD1994}. Unfortunately, this program cannot be carried out for
the extended $1/r$-Hubbard model because the exact ground state is not known
for the effective spin model which is a linear combination
of the Heisenberg model with nearest-neighbor interaction 
and the Haldane-Shastry model with $1/r^2$-exchange interaction~\cite{HaldaneShastryHaldane,HaldaneShastryShastry}.
A variational strong-coupling approach that employs the Baeriswyl wave 
function~\cite{Florianmodelvariational,PhysRevB.51.1993}
can neither be carried out analytically because it requires the evaluation
of $\langle \hat{T}^3\rangle$ in the Gutzwiller-projected Fermi sea.

In the one-dimensional extended Hubbard model, there is a bond-order-wave phase 
below a critical interaction strength $U_{\rm tri}$ that separates the Mott-Hubbard and 
CDW insulators~\cite{Jeckelmann2002,Satoshis2007,MundNoackLegeza2009}. For $U>U_{\rm tri}$, the transition
from the Mott-Hubbard insulator to the CDW insulator is discontinuous.
As can be seen from Fig.~\ref{fig:verticalU17}, we also find indications for the existence
of a bond-order-wave phase. The charge gap of the Mott-Hubbard insulator
closes around $V_{\rm c}(U=1.7)\approx 0.87$ and reopens beyond $V_{\rm c}(U=1.7)$ with a small value. The extrapolation of
the gap remains linear as a function of $1/L$ even at $V=0.88$, as seen in Fig.~\ref{fig:verticalU17}a.
For larger values of $V$, the gap drastically increases and the
extrapolation displays a $1/L^2$ behavior for large~$L$.
The same behavior of the gap was observed for the one-dimensional extended Hubbard model at $U=2W$~\cite{Satoshis2007,MundNoackLegeza2009} where it was numerically shown in detail that
as a function of~$V$ the Mott-Hubbard insulator gives in to a bond-order-wave insulator
before the CDW phase eventually takes over.

Further investigations are necessary to corroborate the
existence of a bond-order-wave phase in the vicinity of the CDW transition
also for the extended $1/r$-Hubbard model.
Note, however that we do not expect a bond-order wave as an intermediate phase for small interactions because
in the extended $1/r$-Hubbard model the metallic Luttinger liquid 
overrides a conceivable bond-order wave.

\section{Conclusions}
\label{sec:conclusions}

In this work we applied the density-matrix renormalization group (DMRG) method
to the half-filled extended $1/r$-Hubbard model where 
the electron transfer amplitudes decay proportional to the inverse
chord distance of two lattice sites on a ring.
The model describes a linear dispersion within the Brillouin zone and thus
provides an ideal case to study the Mott-Hubbard transition
because it lacks Umklapp scattering. Therefore, the
metal-to-insulator transitions occur at finite interaction strengths.
Consequently, all generic phases, namely Luttinger-liquid metal, Mott-Hubbard insulator, and charge-density-wave insulator,
occupy a finite region in the $(U,V)$ ground-state phase diagram, see Fig.~\ref{fig:phasediagram}.

Mapping the quantum phase transition boundaries for the specific model
is one of the main achievements of this work. 
To this end, we use DMRG data for up to $L=80$ sites to calculate 
the ground-state energy, the two-particle gap, the momentum distribution,
the Luttinger parameter, and the structure factor.
The finite-size behavior of the ground-state energy, of the two-particle gap, and of
the structure factor permit to determine the critical interaction parameters
for the instability of the Luttinger liquid metal against the Mott-Hubbard insulator and
the charge-density-wave insulator, respectively. Moreover, we monitor the Luttinger parameter that
also signals the breakdown of the Luttinger liquid metal at a metal-to-insulator transition.
We tested the validity of our analysis against exact results for $V=0$ for which analytic results
for the ground-state energy and the gap exist for all interactions~$U$ and system sizes~$L$.

The phase diagram in Fig.~\ref{fig:phasediagram} shows that the nearest-neighbor interaction and
the Hubbard interaction counteract each other. On the one hand,
the Mott transition 
shifts to larger values, i.e., a weak to moderate nearest-neighbor interaction 
stabilizes the Luttinger liquid metal. Apparently, the two-particle scattering interaction becomes smoother in position space and renders the total interaction less effective.
On the other hand, as can readily be understood from classical considerations,
the Hubbard interaction opposes the formation of a charge-density wave
because, by definition, a CDW augments the particle density on the same lattice site.

In contrast to the `standard' extended Hubbard model in one dimension,
the absence of Umklapp scattering and the competition of both interactions 
leads to an extended metallic region in the phase diagram. 
The extrapolations suggest that there is a tri-critical point where all three phases touch.
It will be interesting to analyze this region in phase space with higher accuracy, i.e.,
more data points in the $(U,V)$ parameter space close to 
$(U_{\rm tri},V_{\rm tri})\approx (1.5,0.75)$, and larger system sizes, $L>80$.
Moreover, a conceivable bond-order wave above the tri-critical point 
between Mott insulator and charge-density-wave insulator should be investigated in more detail.
These tasks are left for future studies.

\begin{acknowledgments}
  \"O.L. has been supported by the Hungarian National Research,
  Development, and Innovation Office (NKFIH) through Grant No.~K134983, and TKP2021-NVA
  by the Quantum Information National Laboratory of Hungary.
  \"O.L. also acknowledges financial support from the Hans Fischer Senior Fellowship program
  funded by the Technical University of Munich -- Institute for Advanced Study and from the Center for
  Scalable and Predictive methods for Excitation and Correlated phenomena
  (SPEC),
  which is funded as part of the Computational Chemical Sciences Program
  by the U.S.\
  Department of Energy (DOE), Office of Science, Office of
  Basic Energy Sciences,
  Division of Chemical Sciences, Geosciences, and Biosciences
  at Pacific Northwest National Laboratory.
\end{acknowledgments}

\appendix

\renewcommand{\thesection}{{\hspace{-3pt}}}
\section{Hartree-Fock theory}
\renewcommand{\thesection}{\Alph{section}}

\subsection{CDW Hartree-Fock Hamiltonian}

In Hartree-Fock theory, we decouple the two-particle interaction as follows,
\begin{eqnarray}
    \hat{D}^{\rm HF}=\hat{D}^{\rm H}&=& \sum_l \langle \hat{n}_{l,\uparrow} \rangle \hat{n}_{l,\downarrow}
+ \hat{n}_{l,\uparrow} \langle \hat{n}_{l,\downarrow} \rangle 
-\langle \hat{n}_{l,\uparrow} \rangle \langle \hat{n}_{l,\downarrow} \rangle \;,\nonumber \\
\mbox{} \\
\hat{V}^{\rm HF} &=& \hat{V}^{\rm H} + \hat{V}^{\rm F} \; ,  \\
\hat{V}^{\rm H} &=& \sum_l \left( \langle \hat{n}_l \rangle -1\right) 
\left( \hat{n}_{l+1}  -1\right)  \nonumber \\[-9pt]
&&\hphantom{\sum_l} + \left( \hat{n}_l -1\right)  \left( \langle \hat{n}_{l+1} \rangle -1\right) 
\nonumber \\
&&\hphantom{\sum_l} - \left( \langle \hat{n}_l \rangle -1\right)  \left( \langle \hat{n}_{l+1} \rangle -1\right) 
\; ,\\
\hat{V}^{\rm F}&=& \sum_{l,\sigma} \langle \hat{c}_{l,\sigma}^+ \hat{c}_{l+1,\sigma}^{\vphantom{+}}\rangle
\hat{c}_{l,\sigma}^{\vphantom{+}} \hat{c}_{l+1,\sigma}^+ \nonumber \\[-9pt] 
&&\hphantom{\sum_{l,\sigma}} + 
\hat{c}_{l,\sigma}^+ \hat{c}_{l+1,\sigma}^{\vphantom{+}}
\langle \hat{c}_{l,\sigma}^{\vphantom{+}} \hat{c}_{l+1,\sigma}^+ \rangle \nonumber\\
&&\hphantom{\sum_l} - 
\langle \hat{c}_{l,\sigma}^+ \hat{c}_{l+1,\sigma}^{\vphantom{+}}\rangle 
\langle \hat{c}_{l,\sigma}^{\vphantom{+}} \hat{c}_{l+1,\sigma}^+ \rangle \; .
\end{eqnarray}
Here, where $\langle \hat{A} \rangle$ denotes the ground-state expectation value 
of the operator $\hat{A}$,
\begin{equation}
    \langle \hat{A}\rangle \equiv \langle \Phi_0 | \hat{A} | \Phi_0\rangle  
\end{equation}
with $|\Phi_0\rangle$ as the ground state of the Hartree-Fock Hamiltonian $\hat{H}^{\rm HF}$, see below.
We make the CDW Ansatz for the order parameter
\begin{equation}
  \langle \hat{n}_{l,\sigma} \rangle   = \frac{1}{2} \left( 1 + (-1)^l \Delta\right)
  \label{appeq:selfconstDelta}
\end{equation}
with the real CDW parameter $\Delta \geq 0$,
and introduce the abbreviation
\begin{equation}
B=   \langle \hat{c}_{l,\sigma}^+ \hat{c}_{l+1,\sigma}^{\vphantom{+}}\rangle = {\rm i}b\; .
\label{appeq:selfconstB}
\end{equation}
Particle-hole symmetry implies that $B$ is purely complex at half band-filling, i.e., $b$ is real.
Note that we disregard a possible bond-order wave (BOW) by assuming that $B$ does not alternate from site to site.

With these abbreviations, we can rewrite the Hartree-Fock interaction at half band-filling as
\begin{eqnarray}
   \hat{D}^{\rm H}&=& \frac{L}{4}\left(1 -\Delta^2 \right)
   +\frac{\Delta}{2} \sum_{l,\sigma} (-1)^l \hat{n}_{l,\sigma} \; , \\
   \hat{V}^{\rm H} &=& L\Delta^2-2\Delta \sum_{l,\sigma} (-1)^l \hat{n}_{l,\sigma} \; , \\
   \hat{V}^{\rm F} &=& 2Lb^2 +{\rm i} b \sum_{l,\sigma}  
   \left[\hat{c}_{l,\sigma}^+ \hat{c}_{l+1,\sigma}^{\vphantom{+}}
   - \hat{c}_{l+1,\sigma}^+ \hat{c}_{l,\sigma}^{\vphantom{+}} \right] .\;
\end{eqnarray}
The resulting single-particle problem defines the Hartree-Fock Hamiltonian for a possible CDW ground state
\begin{equation}
    \hat{H}^{\rm HF}= \hat{T} + U \hat{D}^{\rm H} + V\left( \hat{V}^{\rm H}+\hat{V}^{\rm F}\right) \; .
\end{equation}
It has to be solved self-consistently, i.e., $\Delta$ must be chosen such that the ground state
fulfills Eq.~(\ref{appeq:selfconstDelta}).

\subsection{Diagonalization of the Hartree-Fock Hamiltonian}

In the CDW phase, the Hartree-Fock Hamiltonian is identical for both spin species,
$\hat{H}^{\rm HF}=\sum_{\sigma} \hat{H}_{\sigma}^{\rm HF}$.
Dropping the spin index we must diagonalize 
\begin{eqnarray}
\hat{H}_{\rm sf} &=&\sum_k \epsilon(k) \hat{C}_k^+\hat{C}_k^{\vphantom{+}}
+\left(\frac{U}{2}-2V\right)\Delta \sum_l (-1)^l\hat{n}_{l}  \nonumber \\
&& 
+{\rm i} b \sum_l  
   \left[\hat{c}_l^+ \hat{c}_{l+1}^{\vphantom{+}}
   - \hat{c}_{l+1}^+ \hat{c}_l^{\vphantom{+}} \right] +C
\end{eqnarray}
 for spinless fermions (`sf'), where $C=UL/8(1-\Delta^2)+LV\Delta^2/2+LVb^2$.
 In momentum space, the Hamiltonian reads
\begin{eqnarray}
    \hat{H}_{\rm sf} &=& C+ \sideset{}{'}\sum_k     \Bigl[
    \left(\epsilon(k)+b(k)\right) \hat{C}_k^+\hat{C}_{k}^{\vphantom{+}} \nonumber \\
    && \hphantom{\sideset{}{'}\sum_k     \Bigl[}
    +\left(\epsilon(k+\pi)-b(k)\right) \hat{C}_{k+\pi}^+\hat{C}_{k+\pi}^{\vphantom{+}} \Bigr] \\
    &&+\left(\frac{U}{2}-2V\right)\Delta \sideset{}{'}\sum_k \left(\hat{C}_k^+\hat{C}_{k+\pi}^{\vphantom{+}}
    -\hat{C}_{k+\pi}^+\hat{C}_{k}^{\vphantom{+}}\right)\, ,\nonumber 
\end{eqnarray}
where the prime on the sum indicates the $k$-space region $-\pi<k<0$
and $b(k)= -2bV\sin(k)\geq 0$.

We diagonalize $\hat{H}_{\rm sf}$ with the help of the linear transformation
\begin{eqnarray}
\hat{C}_k &=& c_k \hat{\alpha}_k -s_k \hat{\beta}_k \; , \nonumber \\
\hat{C}_{k+\pi} &=& s_k \hat{\alpha}_k +c_k \hat{\beta}_k  \; ,
\end{eqnarray}
where we abbreviate $c_k \equiv \cos(\varphi_k)$ and $s_k=\sin(\varphi_k)$.
The mixed terms in $\hat{H}_{\rm sf}$ vanish when we demand
\begin{equation}
    \tan(2\varphi_k)= -\frac{(2V-U/2)\Delta}{b(k)+(\epsilon(k)+\epsilon(k+\pi))/2} \geq 0 \;.
\end{equation}
The diagonal terms result in
\begin{equation}
    \hat{H}_{\rm sf} =\sideset{}{'}\sum_k E_{\alpha}(k) \hat{\alpha}_k^+\hat{\alpha}_k^{\vphantom{+}}
    + E_{\beta}(k) \hat{\beta}_k^+\hat{\beta}_k^{\vphantom{+}} +C
\end{equation}
with
\begin{eqnarray}
    E_{\substack{\alpha \\[1pt] \beta}}(k) &=& \frac{1}{2} \left(\epsilon(k) +\epsilon(k+\pi)\right) \mp s(k)\nonumber\\
s(k)&=& \sqrt{\left[2V-\frac{U}{2}\right]^2\Delta^2 +\left[b(k)+\frac{\epsilon(k)-\epsilon(k+\pi)}{2}\right]^2}\nonumber \\
\end{eqnarray}
so that $E_{\alpha}(k)< E_{\beta}(k)$ for all $-\pi<k<0$.
Therefore, the ground state contains only $\alpha$-particles,
\begin{equation}
    |\Phi_0\rangle= \prod_{-\pi<k<0} \hat{\alpha}_{k,\sigma}^+ |{\rm vac}\rangle \; ,
\end{equation}
where we re-introduced the spin index.

\subsection{Self-consistency equations and CDW transition}

The self-consistency equations~(\ref{appeq:selfconstDelta}) and~(\ref{appeq:selfconstB}) become
\begin{eqnarray}
    \Delta &=& \Delta  \int_{-\pi}^0 \frac{{\rm d}k}{\pi} \frac{2V-U/2}{s(k)}\; ,\\
    b&=& -\int_{-\pi}^0 \frac{{\rm d}k}{2\pi} 
    \frac{\sin(k)\left[b(k)+(\epsilon(k)-\epsilon(k+\pi))/2\right] }{s(k)}\nonumber
\end{eqnarray}
in the thermodynamic limit. 
The set \{$\Delta=0$, $b=-1/\pi$\} provides the solution for the Fermi sea of non-interacting particles.

Within Hartree-Fock theory, the CDW transition is continuous. We seek a solution for $\Delta=0^+$
and $b=-1/\pi$ so that $V_{\rm c}(U)$ must obey the equation
\begin{equation}
    \frac{1}{2V_{\rm c}(U)-U/2} = \int_0^{\pi} \frac{{\rm d}k}{\pi} \frac{1}{1/4+2V_{\rm c}(U)\sin(k)/\pi} \;.
\end{equation}
Using {\sc Mathematica}~\cite{Mathematica12} and with the abbreviation $a_{\rm c}=8V_{\rm c}/\pi$ we find
\begin{equation}
     \frac{1}{a_{\rm c}-2U/\pi} = \frac{\pi}{\sqrt{1-a_c^2}}
     -\frac{2}{\sqrt{1-a_c^2}}\arctan\left(\frac{a_c}{\sqrt{1-a_c^2}}\right) \; .
\end{equation}
This equation must be solved numerically for given~$U$. For example, at $U=0$ we find
$a_c\approx 0.394235$ so that $V_{\rm c}(U=0)\approx 0.154816$ in Hartree-Fock theory.
The resulting curve $V_{\rm c}^{\rm HF}(U)$ is shown in Fig.~\ref{fig:phasediagram}.


\bibliography{LRHM}

\end{document}